\documentclass[numberedappendix,twocolumn,onecolappendix]{openjournal}

\usepackage{newtxtext,newtxmath}

\usepackage[T1]{fontenc}
\usepackage{comment}

\DeclareRobustCommand{\VAN}[3]{#2}
\let\VANthebibliography\thebibliography
\def\thebibliography{\DeclareRobustCommand{\VAN}[3]{##3}\VANthebibliography}

\usepackage{graphicx}	
\usepackage{amsmath}	
\usepackage{tikz}
\usepackage{hyperref}

\hypersetup{
    colorlinks=true,
    linkcolor=blue,
    citecolor=blue,
    filecolor=magenta,      
    urlcolor=cyan,
    pdfpagemode=FullScreen,
    }

\newcommand{\Msun}{M_\odot}

\begin{document}

\title{\textsc{Trinity} VII. Predictions for the Observable Correlation Functions of Accreting Black Holes}

\author{Oddisey Knox$^{1}$}
\author{Haowen Zhang$^{1}$}
\author{Peter Behroozi$^{1,2}$}
\affiliation{$^{1}$Department of Astronomy and Steward Observatory, University of Arizona, Tucson, AZ 85721, USA}
\affiliation{$^{2}$Division of Science, National Astronomical Observatory of Japan, 2-21-1 Osawa, Mitaka, Tokyo 181-8588, Japan}



\begin{abstract}
The quasar correlation function  assesses the occurrence of quasar pairs as a function of separation, which is strongly influenced by quasar host halo masses. The empirical \textsc{Trinity} model recently inferred the redshift-dependent relationship between supermassive black hole (SMBH) mass, galaxy mass, and halo mass, using constraints other than correlation functions (e.g., quasar luminosity functions, active galactic nuclei occupation fractions, and SMBH mass--bulge mass relations).  Hence, comparing the predicted quasar correlation functions from \textsc{Trinity} to real observations is an important test of \textsc{Trinity}'s inferred SMBH---halo relation.  In this work, we use a compilation of observed two-point projected and redshift-space correlation functions from $0 \leq z \leq 3.5$.  We find that \textsc{Trinity} accurately predicts quasar correlation functions within observed error bars, although observations do not have much constraining power at lower redshifts due to smaller observable volumes and lower quasar number densities.  This finding is consistent with \textsc{Trinity} having the correct placement of quasars within their host galaxies and dark matter halos, without requiring quasar clustering constraints during model fitting. Using \textsc{Trinity}, we also predict the clustering as a function of quasar bolometric luminosity, finding that existing survey uncertainties are too large to show measurable differences ($\lesssim 0.3$ dex change in bias for $10^{42}$ erg s$^{-1}$ compared to $10^{46}$ erg s$^{-1}$ SMBHs across redshifts).  This fact arises because most SMBH growth (and hence quasar luminosity) occurs in halos in a similar mass range ($10^{12}-10^{13}\Msun$).
\end{abstract}




\section{Introduction}

In the current paradigm of galaxy evolution, a galaxy forms in the center of each dark matter halo, and a supermassive black hole (SMBH) exists at the center of most galaxies \citep[see, e.g.,][for reviews]{Silk12,Somerville15,Naab17,Vogelsberger20}.  Yet, the physical processes by which galaxies and SMBHs grow and influence each others' growth remain uncertain.  On the theoretical side, this has been due to the extreme simulation resolution required to model all relevant processes simultaneously, including gas infall on cosmological scales ($\gtrsim 1$ Mpc) and feedback from stars and the SMBH ($\lesssim$ 0.1 pc).  On the observational side, this has been due to the difficulty in making simultaneous observations of SMBHs in their most significant accretion phase (i.e., the quasar phase) along with their host galaxies, due to the quasar outshining the host galaxy.  In the rare cases where this has been possible, the significant selection effects involved limit the information that can be inferred directly \citep[even recently, e.g.,][]{Li21}.

At the same time, information about quasar host halos and galaxies is available from quasar environments.  More massive halos and galaxies exhibit stronger spatial clustering due to the nature of structure formation in the Lambda Cold Dark Matter ($\Lambda$CDM) paradigm \citep[e.g.,][]{Tinker10,Wechsler2018}.  Hence, even when the host properties of quasars are not directly visible, their autocorrelation functions and cross-correlation functions with galaxies yield information on their host galaxies and dark matter halos.  The most straightforward interpretation of the clustering yields both a characteristic host halo mass (typically $M_h \sim 10^{12}-10^{13}$ $\Msun$, independent of quasar luminosity or redshift) and an inferred duty cycle, which is typically $\lesssim 0.1\%$, though this is highly dependent on sample selection \citep[e.g.,][]{Croom05,daAngela08,Ross09,Shen2009data,Eftekharzadeh15,Chehade16,Laurent17}.

Many models of SMBH evolution have hence used quasar correlation functions to constrain SMBH occupation in galaxies \citep[e.g.,][]{Hopkins2008,Croton2009,Shen2009model,Conroy2013,Shankar2020Nat}.  The lack of luminosity dependence in quasar clustering most simply implies large scatter in the luminosity distribution of SMBHs, which is consistent with observations \citep[e.g.,][]{Bongiorno2012,Aird2018}.  In addition, the lack of redshift dependence in the host halo mass most simply implies that most SMBH growth occurs in halos with $M_h \sim 10^{12}-10^{13}$ $\Msun$, which is also the mass range hosting most galaxy growth \citep[e.g.,][]{Moster18,Behroozi2019}.

We recently introduced an empirical model, \textsc{Trinity}, which uses an alternate approach to infer the halo--galaxy--SMBH relationship from observations \citep{Zhang2023}. In \textsc{Trinity}, the halo--galaxy relationship is constrained primarily through observed galaxy number densities \cite[see][for a review]{Wechsler2018}, and halo growth histories in dark matter simulations then constrain galaxy growth histories.  The galaxy--SMBH relationship is then primarily constrained by the $z=0$ SMBH--galaxy relationship (specifically, the SMBH mass--galaxy bulge mass relationship; e.g., \citealt{Haring2004,Kormendy2013,McConnell2013}) and the AGN luminosity distributions of galaxy progenitors at higher redshifts \citep[as measured in][]{Aird2018}.  In practice, this gives similar information as applying the So\l{}tan argument to SMBHs in bins of $z=0$ host galaxy stellar mass, summing AGN luminosity along galaxy growth histories to infer SMBH growth histories \cite[see also][]{Shankar2020MNRAS}.

In this paper, we apply the best-fit halo--galaxy--SMBH relationship from \textsc{Trinity} to halos in a dark matter simulation and measure the resulting quasar autocorrelation functions.  Of note, although \textsc{Trinity} combined many different data types to constrain SMBH and galaxy growth, it did not incorporate any quasar clustering data.  Hence, the results here represent a pure prediction of the model.  A match between \textsc{Trinity}'s predictions and observations would suggest that the autocorrelation functions do not provide significant additional information about the SMBH--galaxy--halo relationship beyond that already in observed AGN luminosity distributions of galaxies.  On the other hand, a mismatch would suggest that quasars and galaxies have a more complicated relationship, and that quasar correlation functions are key to interpreting it.  We note that a similar test was performed in \cite{Aird2021}. In that paper, the authors compared to measured quasar biases; here, we compare to the full 2D projected and 3D redshift-space clustering as a function of distance scale.  We also seek to understand, using predictions from \textsc{Trinity}, where the observed lack of luminosity dependence will begin to break down, and hence where quasar selection criteria will more strongly influence clustering \citep [see also][]{Powell2024}.

In Section \ref{s:methods}, we introduce the dark matter simulation used and relevant information about \textsc{Trinity} and the auto-correlation function calculation.  In Section \ref{s:observations}, we describe the observations to which we compare our predictions.  In Section \ref{s:results}, we show the comparisons between \textsc{Trinity}'s predictions and the observations.  We discuss these results in Section \ref{s:discussion} and conclude in Section \ref{s:conclusions}.  Throughout this paper, we assume a flat, $\Lambda$CDM cosmology with parameters $h= 0.67$, $\Omega_\Lambda = 0.693$, $\Omega_m= 0.307$, $n= 0.96$, and $\sigma_8= 0.823$, consistent with the cosmology constraints in \cite{Planck2016}.

\section{Methods}

\label{s:methods}

\subsection{Dark Matter Simulations}
For this work, we use the \textit{MultiDark Planck 2} simulation (\textit{MDPL2}; \citealt{Klypin16, Rodriguez-Puebla16})  with a box size of 1000 comoving Mpc $h^{-1}$, 3840$^3$ particles, a mass resolution of $1.51 \times 10^9 M_\odot h^{-1}$, and a force resolution of 13 kpc $h^{-1}$ at high redshifts and 5 kpc $h^{-1}$ at low redshifts.  The simulation was evolved from $z = 120$ until the present, and assumed a flat $\Lambda$CDM cosmology ($h= 0.67$, $\Omega_\Lambda = 0.693$, $\Omega_m= 0.307$, $n= 0.96$, and $\sigma_8= 0.823$). Halos were found using the \textsc{Rockstar} halo finder \citep{Behroozi13a} and merger trees were constructed using the \textsc{Consistent Trees} code \citep{Behroozi13b}.  This simulation resolves all the halos expected to be hosting luminous quasars, while at the same time covering a large enough volume to minimize sample variance and have accurate large-scale clustering.

    \begin{table*}
    \centering
    \caption{Observational data sets}
    \label{tab:observations}
    \begin{tabular}{ccccccc}
    	\hline
    	Reference & 2D/3D & Redshift Range & Selection & Survey Name & Area (deg$^2$) & Quasar Density (deg$^{-2})$\\
    	\hline
    	\cite{Croom05}& 3D & 0.3 - 2.48 & Optical & 2QZ & 721.6 & 31\\
    	\cite{daAngela08} & 3D & 0.3-2.9 & Optical & 2SLAQ & 180 & 35\\
    	\cite{Ross09} & 2D+3D & 0.3 - 2.1 & Optical & SDSS DR5Q & 4000 & 7.5\\
    	\cite{Eftekharzadeh15} & 2D+3D & 2.2 - 3.4 & Optical & SDSS III- BOSS & 6950 & 10.6\\
    	\cite{Chehade16} & 3D & 0.8 - 2.5 & Optical & 2QDESp, 2SLAQ, 2QZ, SDSS DR5 & 150-4000 & 8-67\\
    	\cite{Laurent17} & 3D & 0.9 - 2.2 & Optical & SDSS IV- eBOSS & 1200 & 57\\
    	\hline
    	\end{tabular}
    \end{table*}

\subsection{The \textsc{Trinity} empirical model for supermassive black holes}

\label{s:trinity}


\textsc{Trinity} is a self-consistent, empirical model that infers the dark matter halo--galaxy--SMBH connection \citep{Zhang2023}. \textsc{Trinity} begins with the distribution of dark matter halo masses as a function of redshift from an N-body simulation. From here, it parameterizes the scaling relations between halo, galaxy and SMBH masses, as well as the shape of the SMBH accretion rate distribution.  Each point in this parameter space fully specifies a unique recipe to populate halos with galaxies and SMBHs as functions of the host halo's mass and redshift. For each trial point in parameter space, \textsc{Trinity} then predicts galaxy and SMBH properties in a mock universe and compares them with real observations. For galaxies, these observations include stellar mass functions, specific star formation rates, cosmic star formation rates, fractions of quiescent galaxies, and $z \geq 8$ UV luminosity functions. These data collectively cover a redshift range of $0 \leq z \leq 13$, including the latest $9 \leq z \leq 13$ galaxy UV luminosity functions from \citet{Harikane2023}. For supermassive black holes, these  observations include the $z=0$ SMBH-bulge mass relation, active black hole mass functions, the quasar luminosity function, and active galactic nuclei (AGN) occupation fractions for all available redshifts in the range $0 \leq z \leq 6.5$. The difference between the predictions and observations determines the likelihood for each point in parameter space, and a Markov Chain Monte Carlo algorithm is used to generate new points in parameter space to explore.  This approach results in the posterior distribution of SMBH--galaxy--halo relationships that match all observed data; in this study, we use the best-fitting relationship.  As mentioned previously, no correlation function data constrained \textsc{Trinity}, so the results in this paper represent a pure model prediction. 

\subsection{Computing quasar correlation functions}

We apply \textsc{Trinity} to the \textit{MDPL2} simulation to generate the mock galaxy and SMBH catalogs used in this work. When constructing these catalogs, we determined the probabilities of galaxies hosting actively accreting BHs above a certain luminosity threshold based on the SMBH luminosity distribution predicted by \textsc{Trinity}: 
\begin{equation}
    f_{>L_\mathrm{thresh}}(M_h,z)= \int_{L_\mathrm{thresh}}^\infty P(L | M_h,z)  \,dL \ ,
\end{equation}
where M$_{h}$ is peak historical halo (or subhalo) mass and $P$ is the luminosity distribution at a fixed halo mass and redshift from the best-fitting \textsc{Trinity} model.
The luminosity distributions $P(L | M_h,z)$  from \textsc{Trinity} are consistent with the AGN observational data listed in Section \ref{s:trinity}, and are primarily constrained by the AGN occupation fraction in galaxies (giving $P(L|M_\ast,z)$) as well as galaxy number densities and correlation functions (giving $P(M_\ast|M_h,z)$). We then weight each halo and subhalo by $f_{>L_\mathrm{thresh}}(M_h,z)$ when calculating its contribution to the quasar two-point correlation function (2PCF).  Intuitively, a longer fraction of time spent above the chosen luminosity threshold corresponds to a higher weight, and vice versa for a shorter fraction of the time (see also \citealt{Aird2021} for a similar approach).  Because we compute quasar clustering based on the weights for all galaxies that may host a quasar (instead of those that stochastically light up as quasars at a single point in time), we can predict quasar clustering down to smaller scales ($\sim$0.2 Mpc) than measurable in most observational samples. 

We compute the projected 2PCF by integrating the standard 3D correlation function: 
\begin{equation}
    w_p(R_p)= \int_{-\pi_\mathrm{max}}^{\pi_\mathrm{max}} \xi(R_p, \pi) d\pi .
\end{equation}
Here, $\pi$ is the line-of-sight distance including redshift-space distortions from peculiar motion, and {$R_p$} is the projected radius. The maximum line-of-sight distance, $\pi_\mathrm{max}$, was set to half the simulation box size, as there is relatively little difference in $w_p$ for values of $\pi_\mathrm{max}>60$ Mpc $h^{-1}$.  The redshift-space 2PCF, $\xi(s)$, is also calculated in the standard way, by averaging $\xi(R_p,\pi)$ over the circle with $R_p^2 + \pi^2 = s^2$.

\section{Observations}
\label{s:observations}

The observational data used in this paper are taken from the 2QZ \citep{Smith05}, 2SLAQ \citep{Richards05}, 2QDESp \citep{Shanks15}, and SDSS  surveys \citep{York00} (including DR5Q \citealt{Schneider07}), SDSS III-BOSS \citep{Eisenstein11}, and SDSS IV-eBOSS \citep{Dawson16}. Observationally, it is difficult to find pairs of quasars on small scales due to the low number density of sources, so most observational constraints on autocorrelation functions are limited to scales greater than $\sim 1-2$ Mpc.  See Table \ref{tab:observations} for full details of the observational data sets, including detection method, redshift ranges, and sky areas.  Typical luminosities of quasars in these observational samples are $L_\mathrm{bol}>10^{46}$ ergs s$^{-1}$ \citep{Ross09}, with very little variation in quasar clustering with luminosity \citep[e.g.,][]{Eftekharzadeh15}, and so we adopt $L_\mathrm{thresh}=10^{46}$ ergs s$^{-1}$ for comparison with \textsc{Trinity}.  When a given reference provides only luminosity-binned clustering without presenting a combined sample, we show the luminosity bin with least uncertainties in the main text, and show each luminosity bin individually in Appendix \ref{a:luminosities}.

\section{Results}

\label{s:results}

\begin{figure*}
    \begin{tabular}{ll}
	\includegraphics[width =\columnwidth]{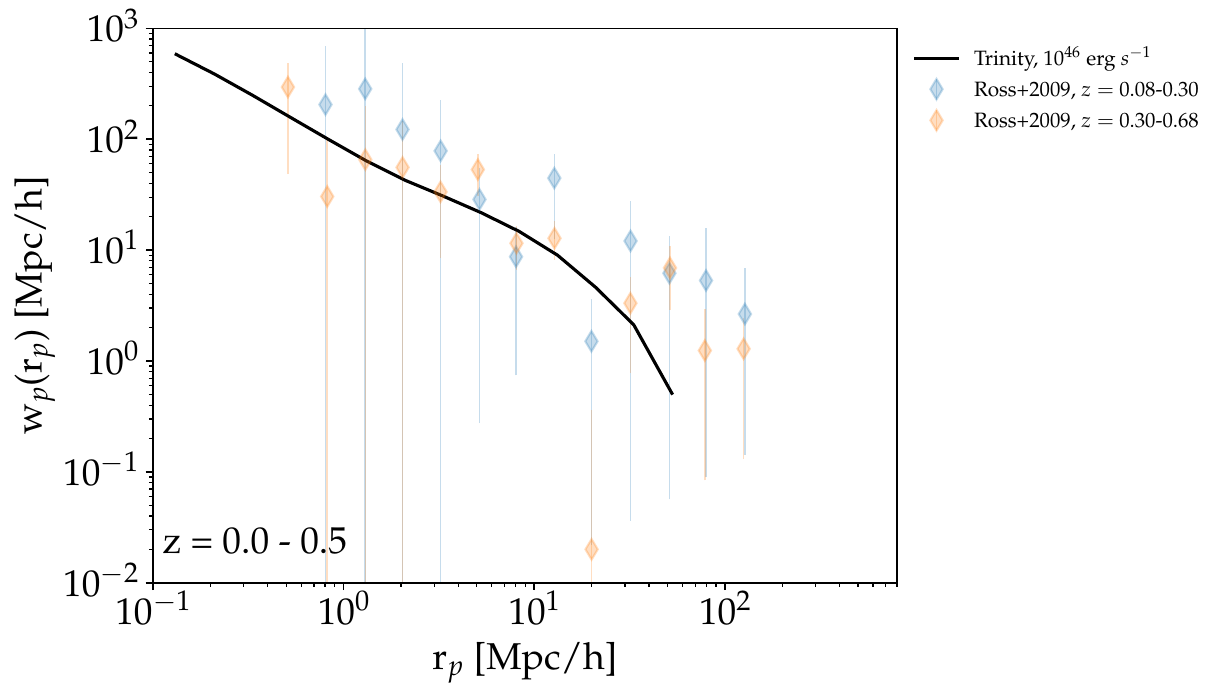} &
    \includegraphics[width =\columnwidth]{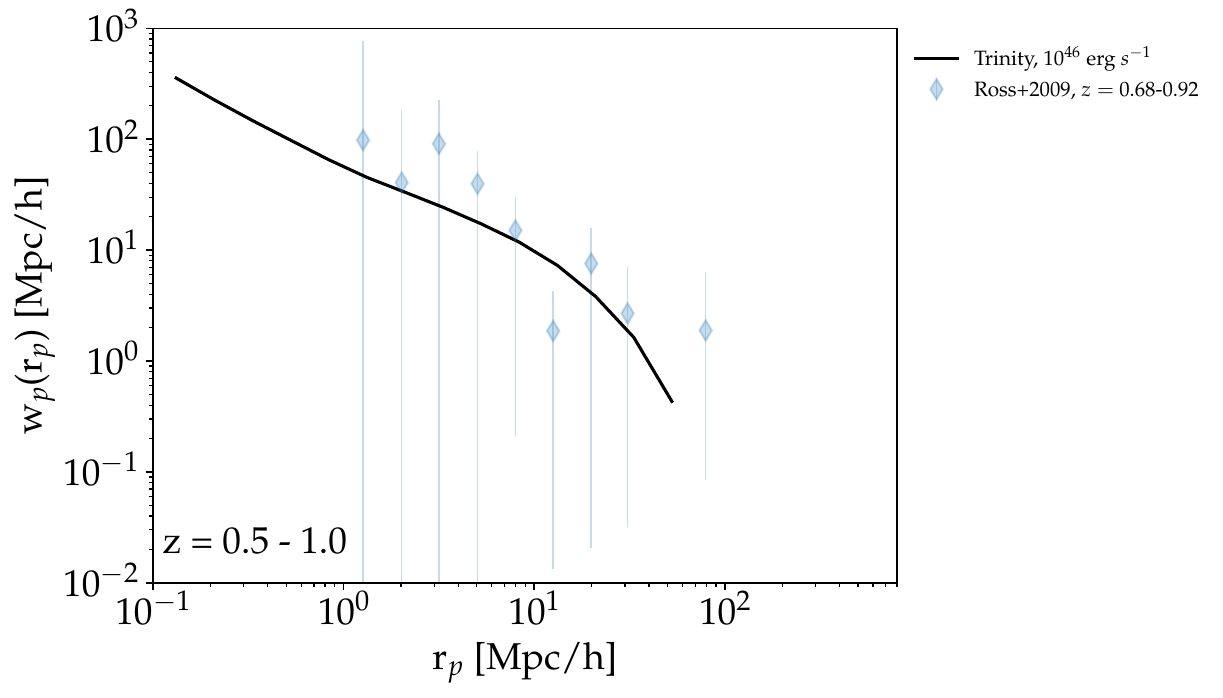}\\
    \includegraphics[width =\columnwidth]{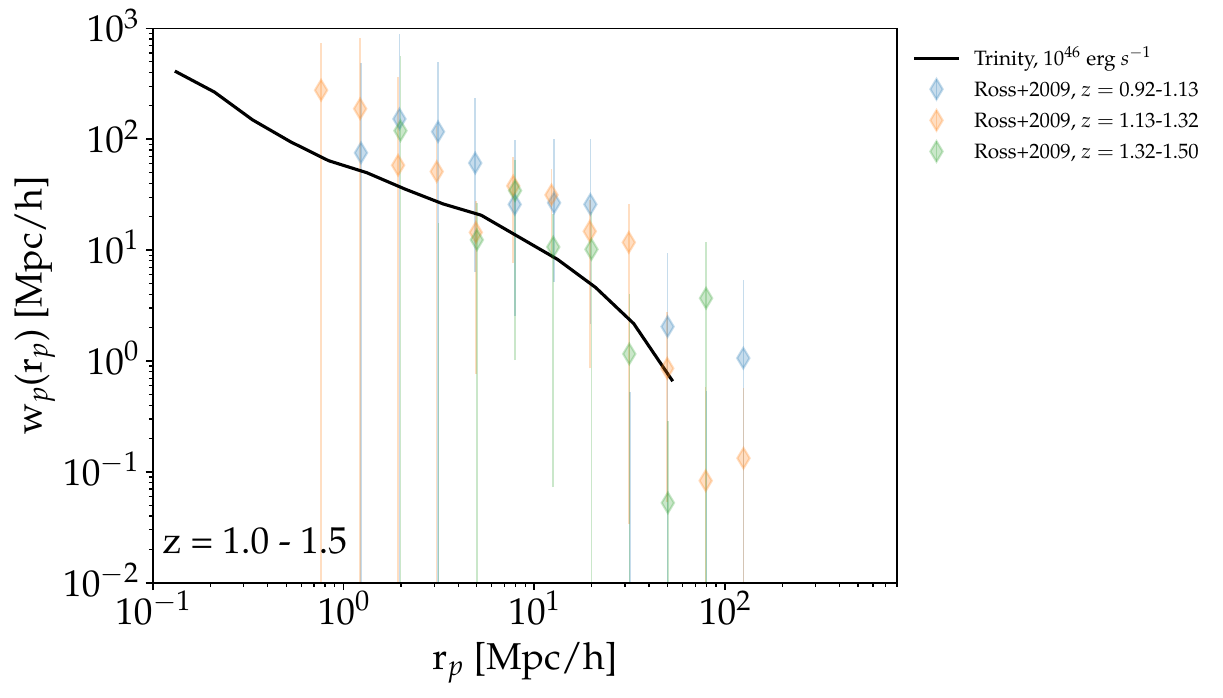} &
    \includegraphics[width =\columnwidth]{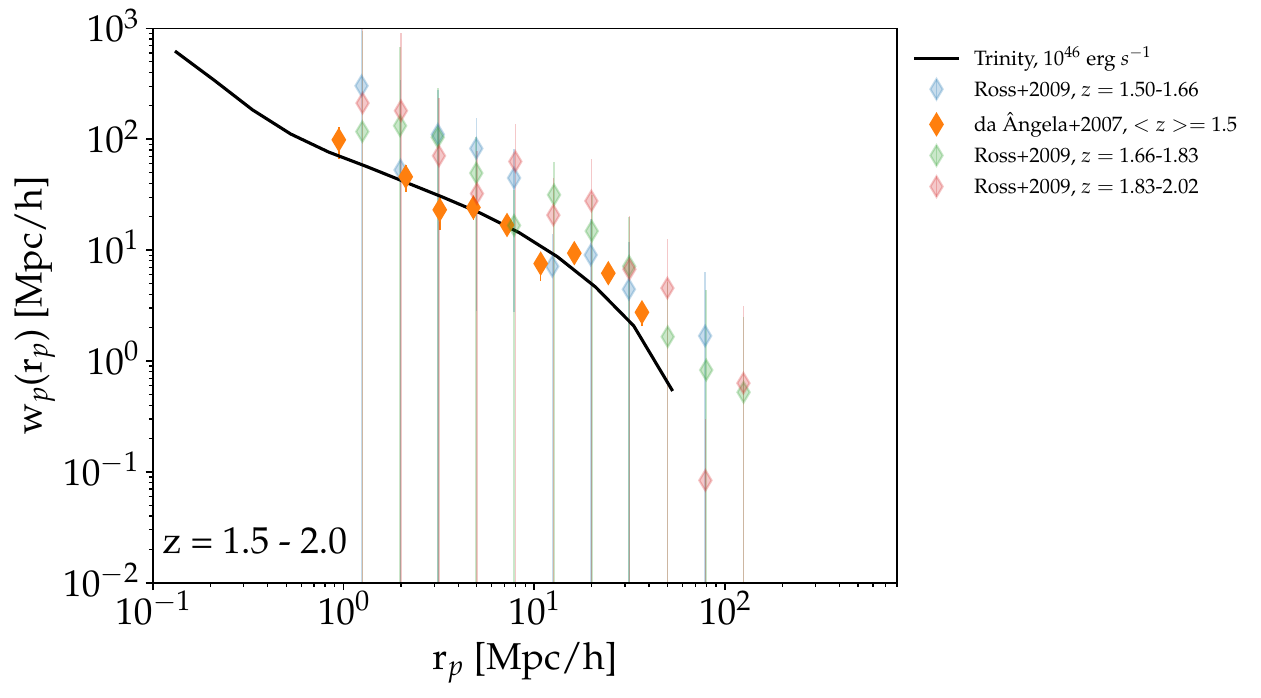}
\end{tabular}    
    \caption{The two-point projected correlation function, $w_p$, for $z = 0-2$ in $\Delta z= 0.5$ bins.  \textsc{Trinity} predictions are shown as the black solid line, with observations shown as symbols. Observations with $>2$ dex uncertainties are shown in transparent hues so that more precise measurements stand out visually.}
    \label{fig:2d}
\end{figure*}

\begin{figure*}
    \begin{tabular}{ll}
    \includegraphics[width = \columnwidth]{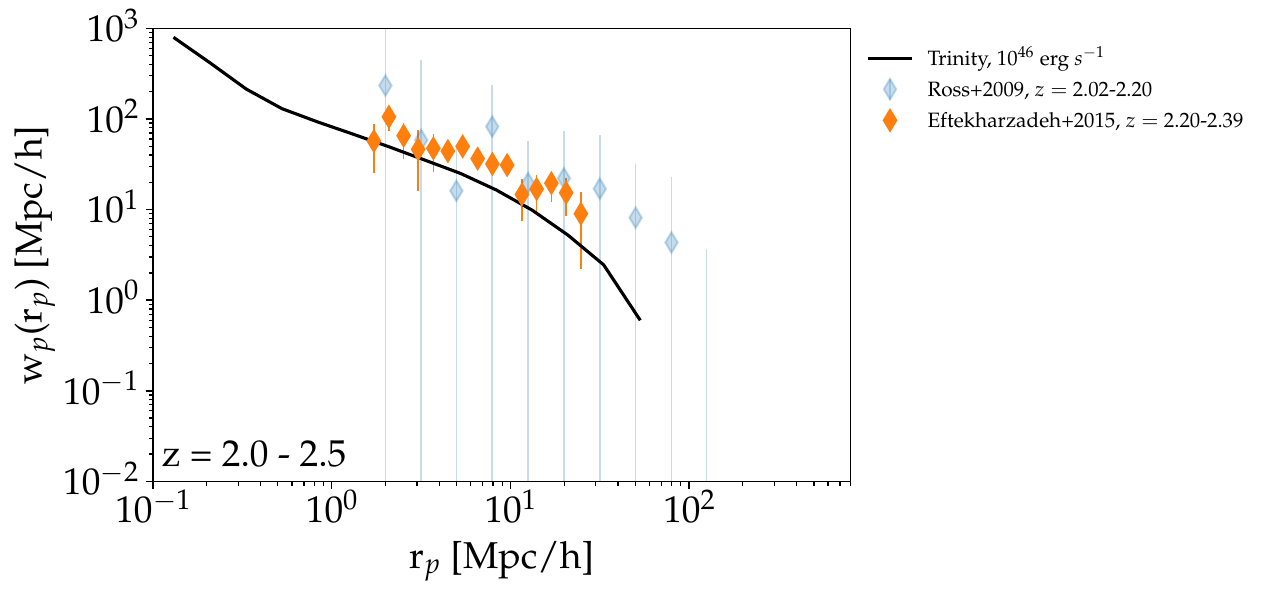} & \hspace{-0.5cm}
    \includegraphics[width = \columnwidth]{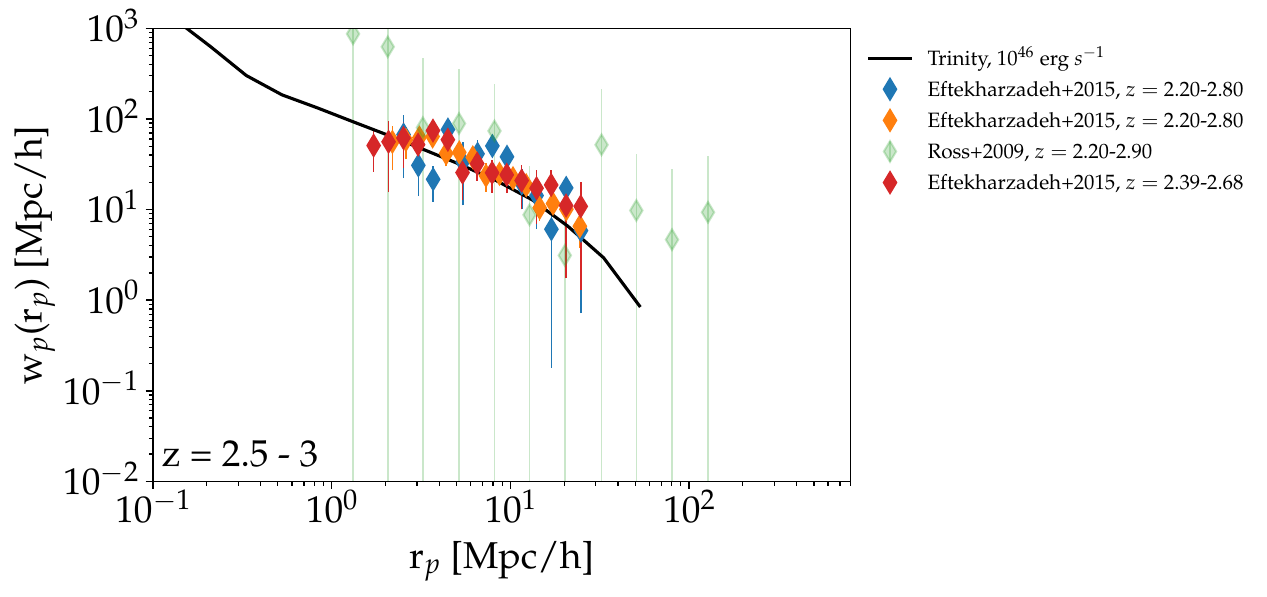}\\
    \multicolumn{2}{c}{\includegraphics[width = \columnwidth]{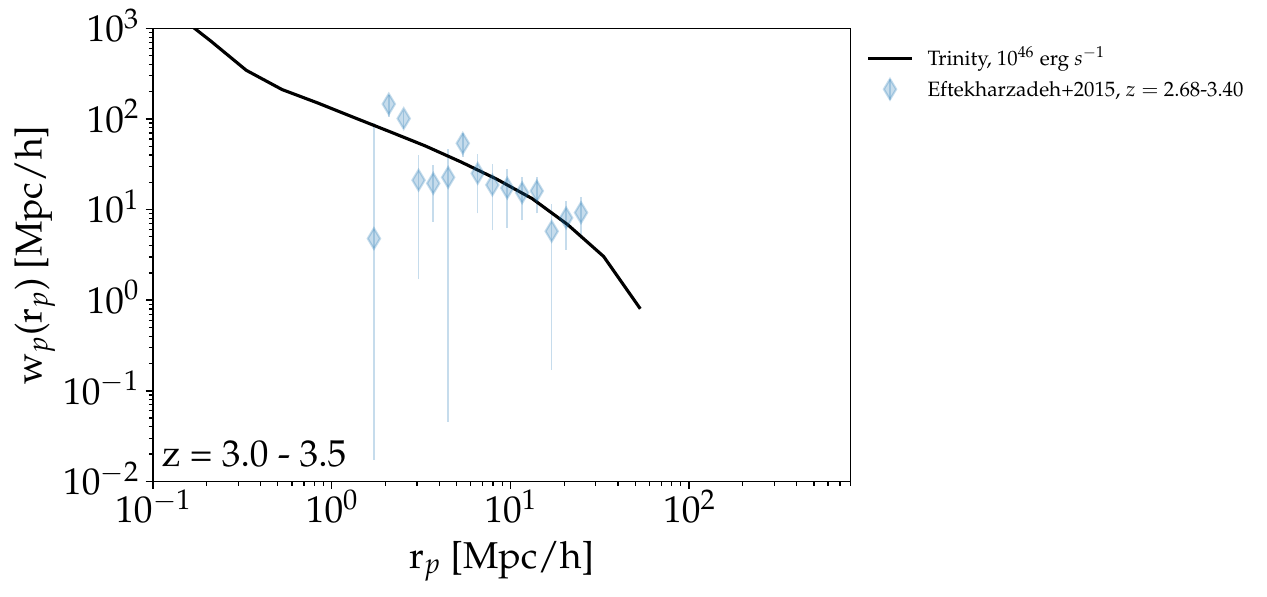}}
    \end{tabular}
    \caption{The two-point projected correlation function, $w_p$, for $z = 2-3.5$ in $\Delta z= 0.5$ bins. \textsc{Trinity} predictions are shown as the black solid line, with observations shown as symbols.  Observations with $>2$ dex uncertainties are shown in lighter colors so that more precise measurements stand out visually.}
    \label{fig:2d_2}
\end{figure*}

\begin{figure*}
	\begin{tabular}{ll}
    \includegraphics[width = \columnwidth]{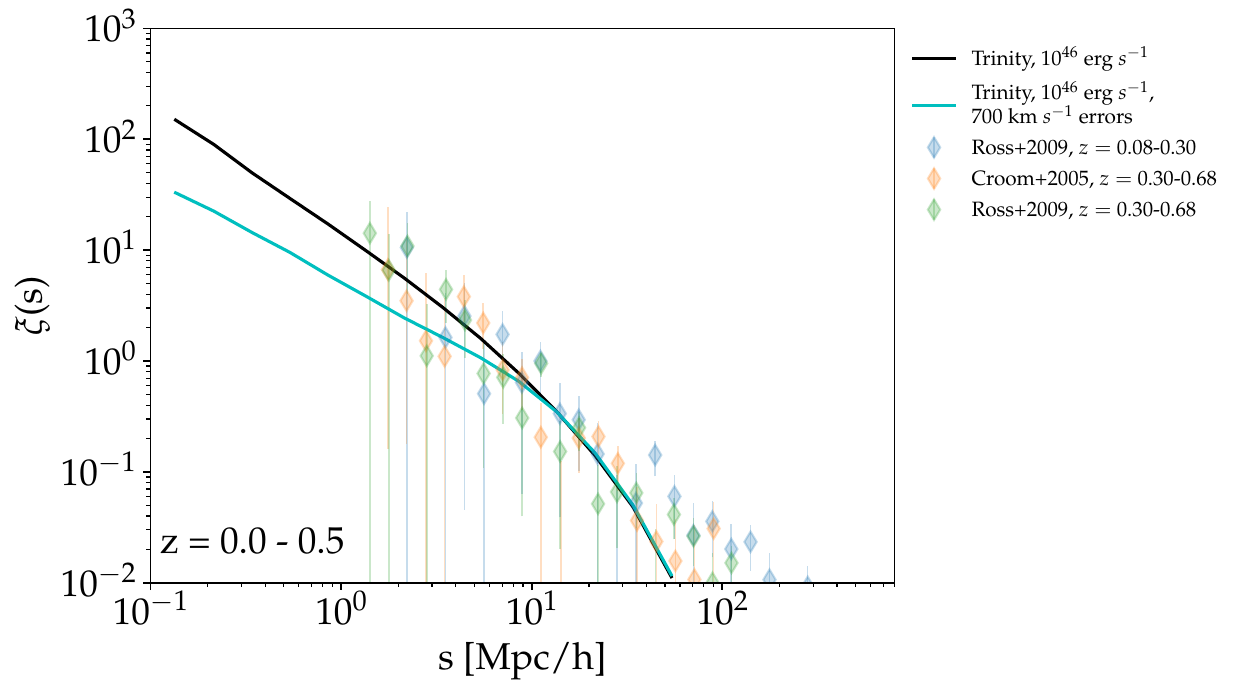} &
    \includegraphics[width = \columnwidth]{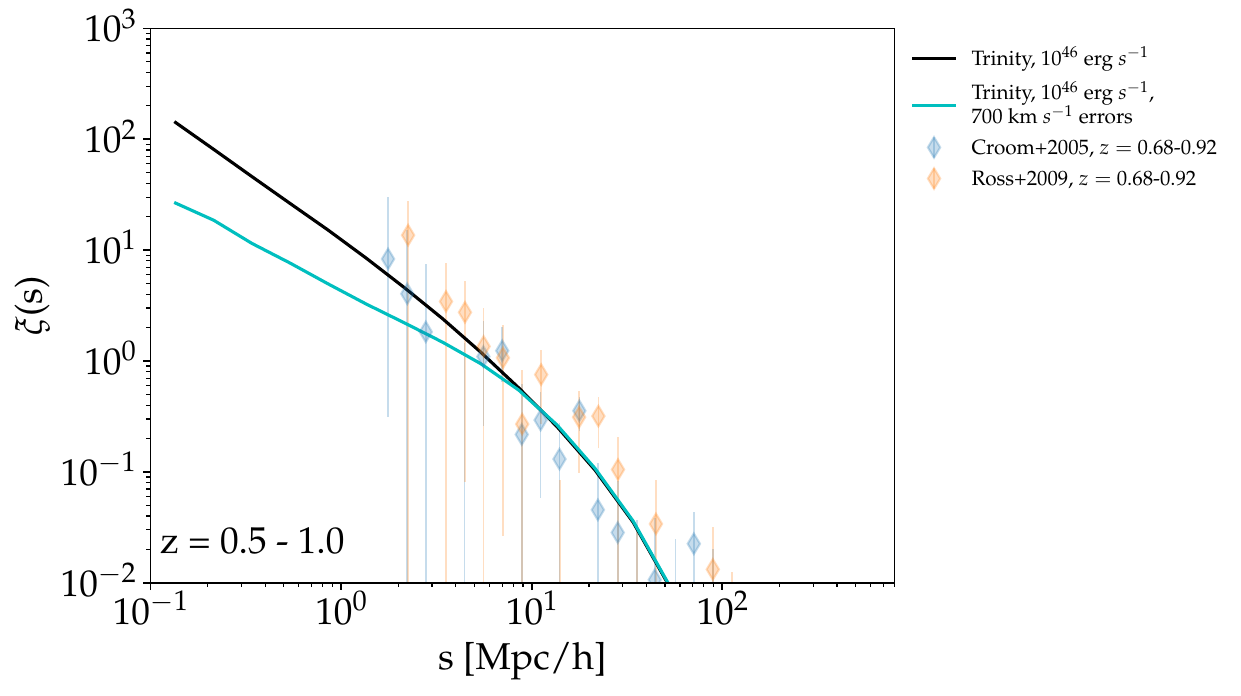}\\ 
    \includegraphics[width = \columnwidth]{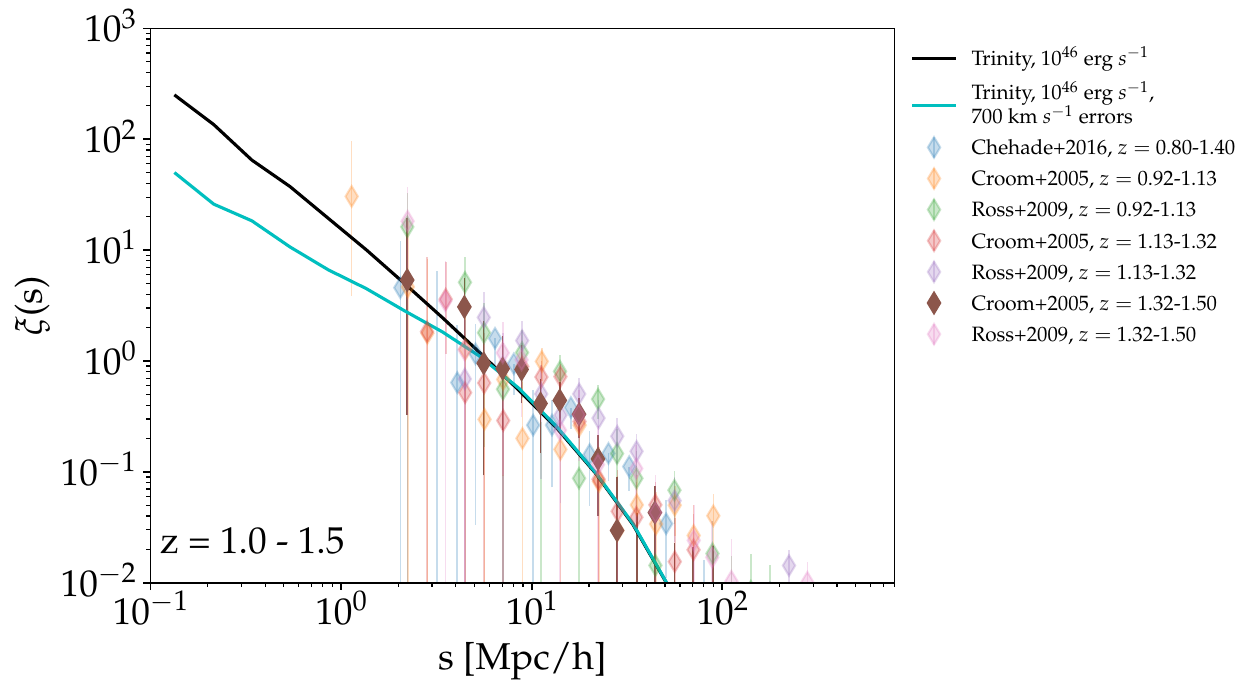} &
    \includegraphics[width = \columnwidth]{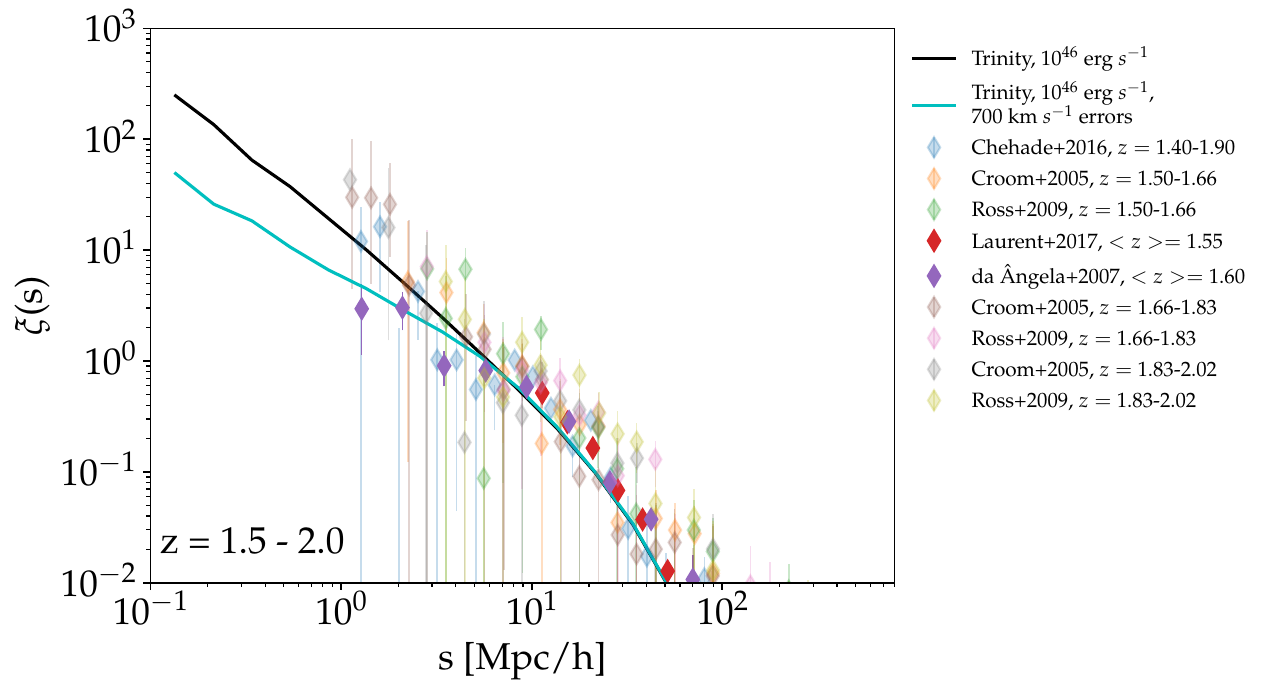}
    \end{tabular}
    \caption{The 3D redshift-space correlation function, $\xi_s$, for $z = 0-2$ in $\Delta z= 0.5$ bins. \textsc{Trinity} predictions are shown as the black solid line (no redshift-space errors) and the blue solid line (700 km s$^{-1}$ redshift-space errors), with observations shown as symbols.  Observations with $>2$ dex uncertainties are shown in lighter colors so that more precise measurements stand out visually.}
    \label{fig:3d}
\end{figure*}

\begin{figure*}
	\begin{tabular}{ll}
	\hspace{-0.5cm}\includegraphics[width = \columnwidth]{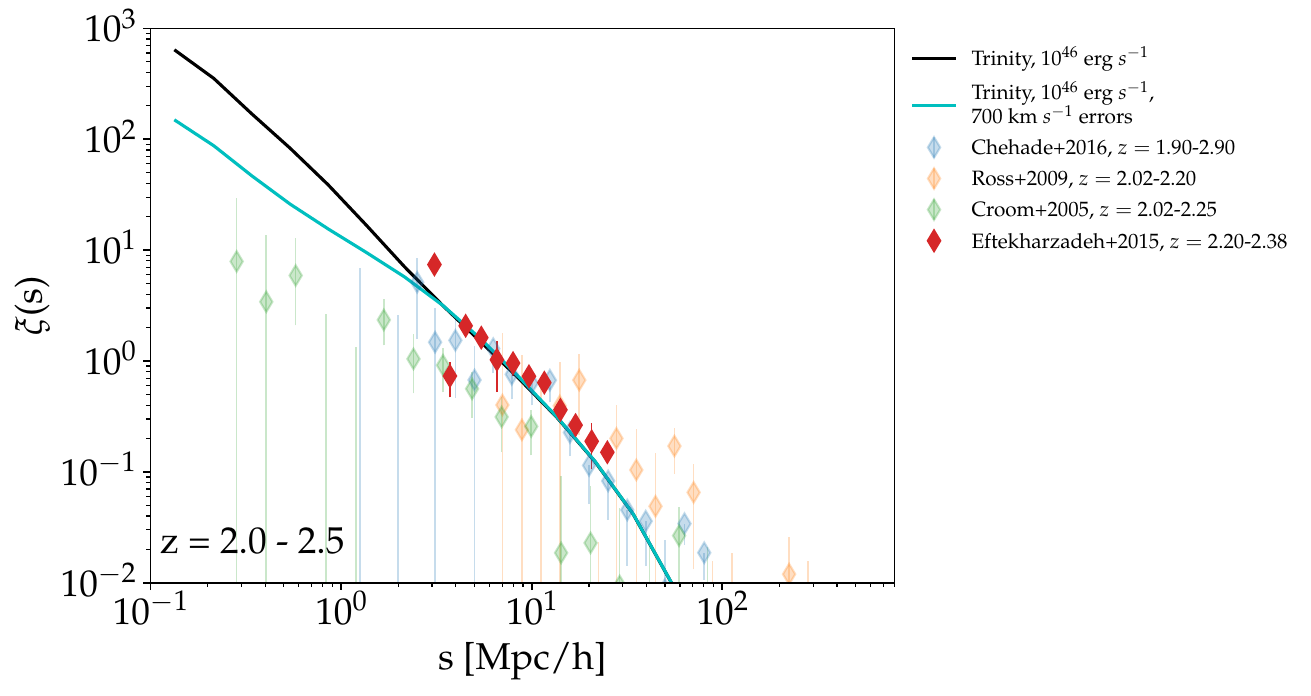} & \hspace{-0.5cm}   \includegraphics[width = \columnwidth]{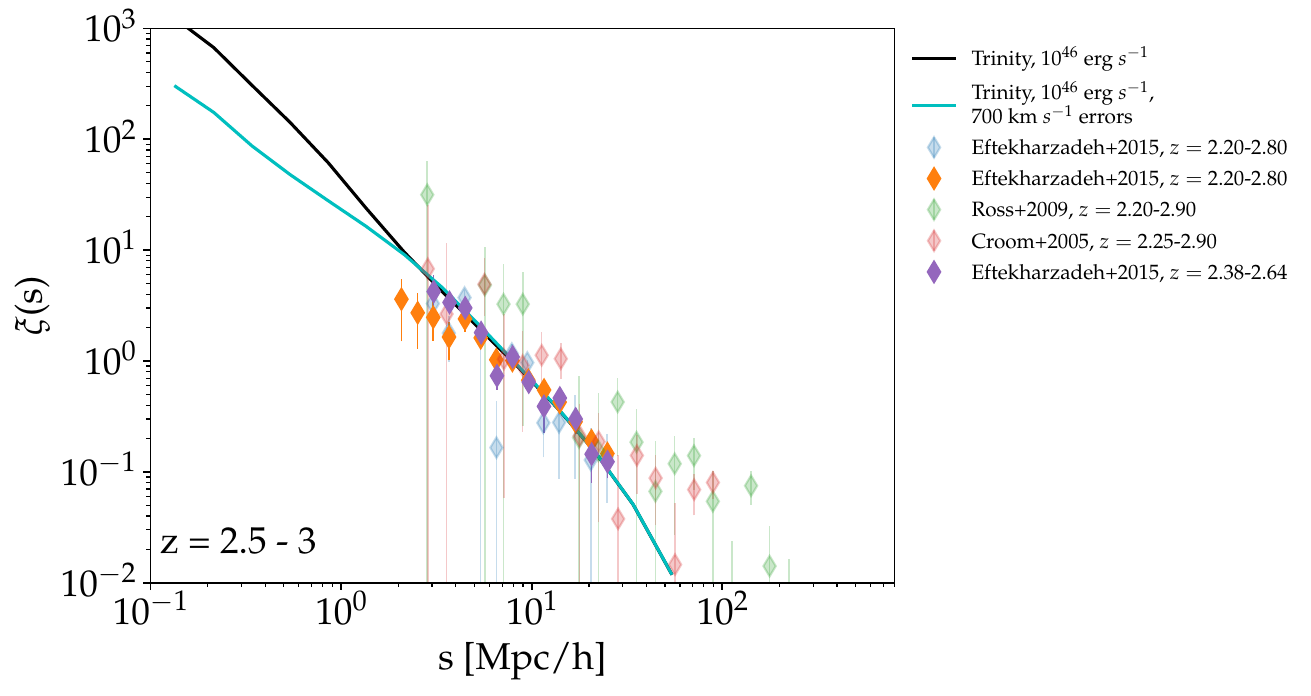}\\
	\multicolumn{2}{c}{\includegraphics[width = \columnwidth]{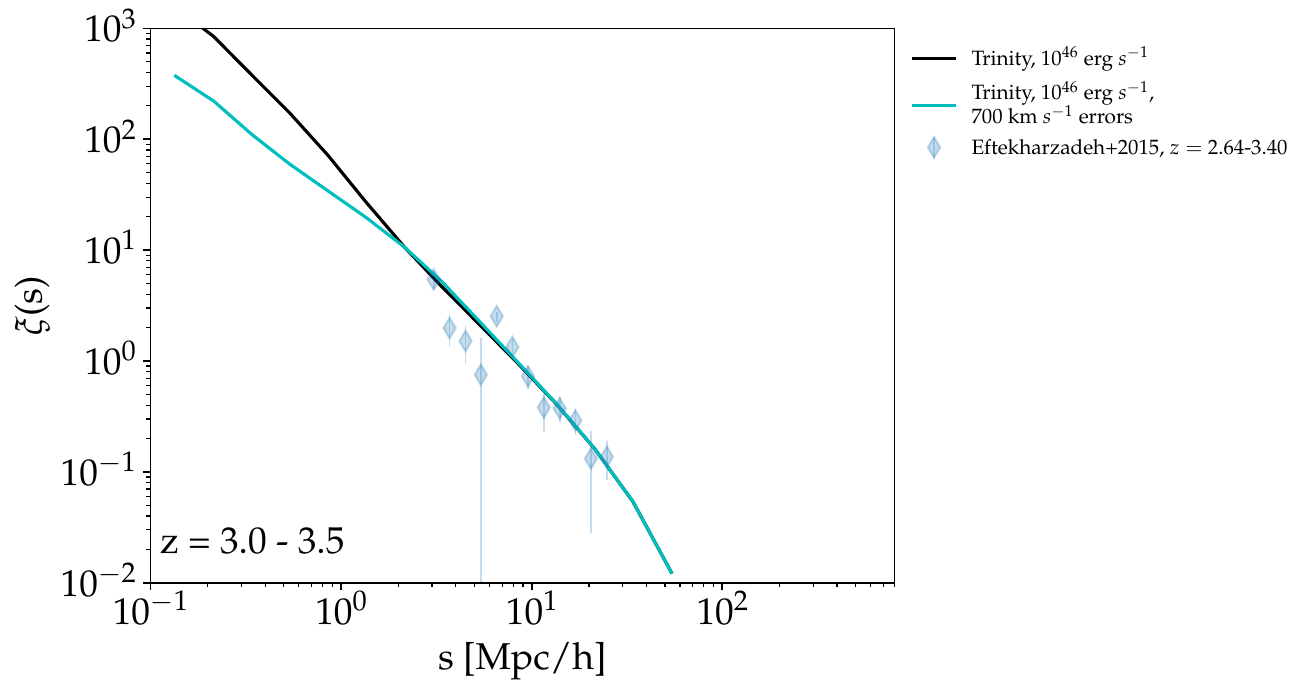}}  
	\end{tabular}
    \caption{The 3D redshift-space correlation function, $\xi_s$, for $z = 2-3.5$ in $\Delta z= 0.5$ bins. \textsc{Trinity} predictions are shown as the black solid line (no redshift-space errors) and the blue solid line (700 km s$^{-1}$ redshift-space errors), with observations shown as symbols.  Observations with $>2$ dex uncertainties are shown in lighter colors so that more precise measurements stand out visually.}
    \label{fig:3d_2}
\end{figure*}


    The projected two-point correlation functions for the redshift range $0 < z < 3.5$ in bins of $\Delta z=0.5$ are given in Figures \ref{fig:2d} and \ref{fig:2d_2}. At $z<1.5$, there is substantial uncertainty in the observed data points due to both smaller observable volumes and lower typical quasar luminosities, making the sample sizes much smaller. Constraints on the correlation function at and below this redshift prove more difficult than at higher redshifts, where quasars are more numerous. Therefore, all data points with uncertainties larger than 2 dex are shown with transparent hues in the figures.  At higher redshifts, the data are more precise, as seen with \cite{daAngela08} at $\langle z \rangle =1.5$ and \cite{Eftekharzadeh15} at $z=2.20-2.80$ and $z=2.68$ from 1-10 Mpc.  Overall, \textsc{Trinity}'s predictions agree well with observations within the uncertainties.

    On large scales ($r_p\gg 10$ Mpc $h^{-1}$), there are known issues with calculating correlation functions that affect \citet{Eftekharzadeh15}, driven by difficulties in accounting for observational systematics at the required precision (including large-scale variations in number densities and survey depth).   Per that author's suggestion (S.\ Eftekharzadeh, priv.\ comm.), we have limited the comparison to $r_p<25$ Mpc/h.  \textsc{Trinity} predicts a downturn in the quasar autocorrelation function at similar scales ($r_p>25$ Mpc/h) which is not an artifact of the model or box size; instead, this feature arises from the shape of the matter power spectrum and is universally found in N-body simulations regardless of size \citep[e.g.,][]{Klypin16}.

    The two-point redshift-space correlation functions are given in Figures \ref{fig:3d} and \ref{fig:3d_2}.  Many of the conclusions above also apply to redshift-space clustering. For example, at $z<1.5$, there are few precise measurements of the quasar correlation function with which we can compare.  At higher redshifts, $z \ge 2$, we see good agreement with both \cite{daAngela08} and \cite{Eftekharzadeh15}. {In the case of \cite{Eftekharzadeh15}, BOSS reaches two magnitudes deeper than SDSS, resulting in 15 times as many quasars at $z \sim 2.5$.}  \cite{daAngela08}, though covering a much smaller area, $\sim$180 deg$^2$, reaches much fainter magnitudes of $g=21.85$ ($i \approx 21.45$), resulting in 8500 quasars. We see minor differences in downturns in the 3D correlation function of the observational data at low redshift-space separations, e.g., $z \lesssim 2.0$ at $\lesssim 5$ Mpc, and at slightly higher redshifts of $z= 2.0-3.0$ from $\sim 2-3$ Mpc. As this may be in part due to redshift uncertainties and systematic quasar--galaxy velocity offsets, we have tested adding a random scatter of 700 km s$^{-1}$ in the redshift-space direction when computing $\xi(s)$ (the cyan curves in Figure \ref{fig:3d} and Figure \ref{fig:3d_2}); this is similar to previous estimates (e.g., 690 km s$^{-1}$ in \citealt{Croom05}). We note that observational color cuts are not part of the selection criteria for \textsc{Trinity}, which will inevitably result in some of the differences observed.

\section{Discussion}

\begin{figure*}
	\includegraphics[width = \columnwidth]{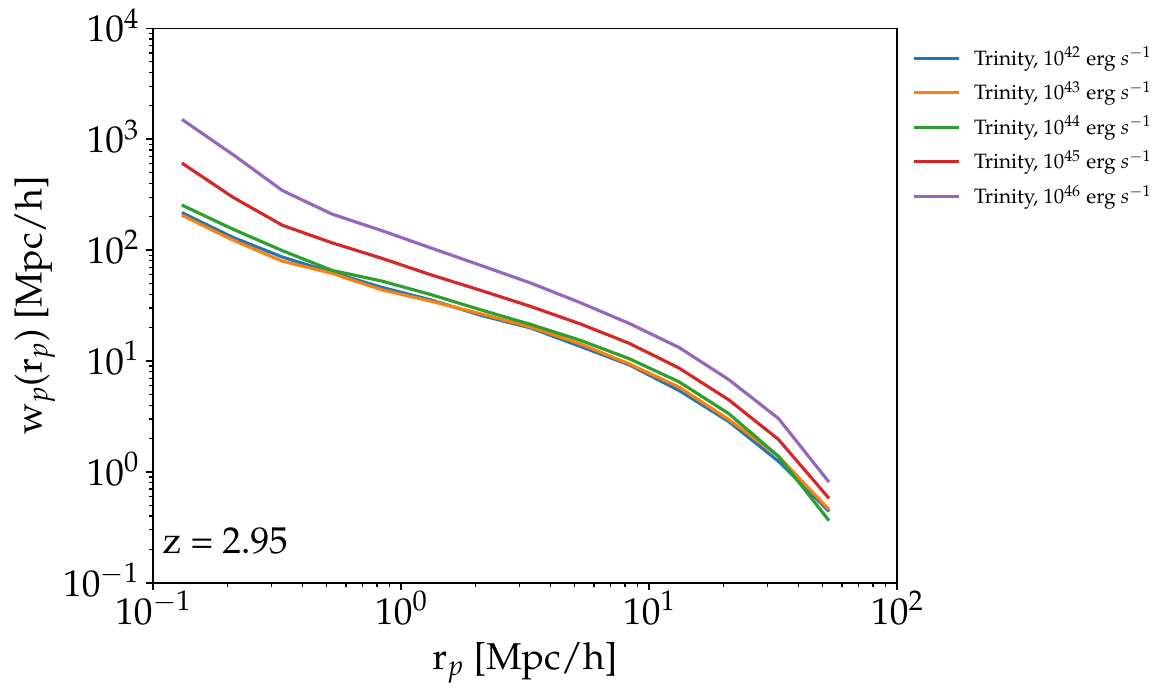} 
 	\includegraphics[width = \columnwidth]{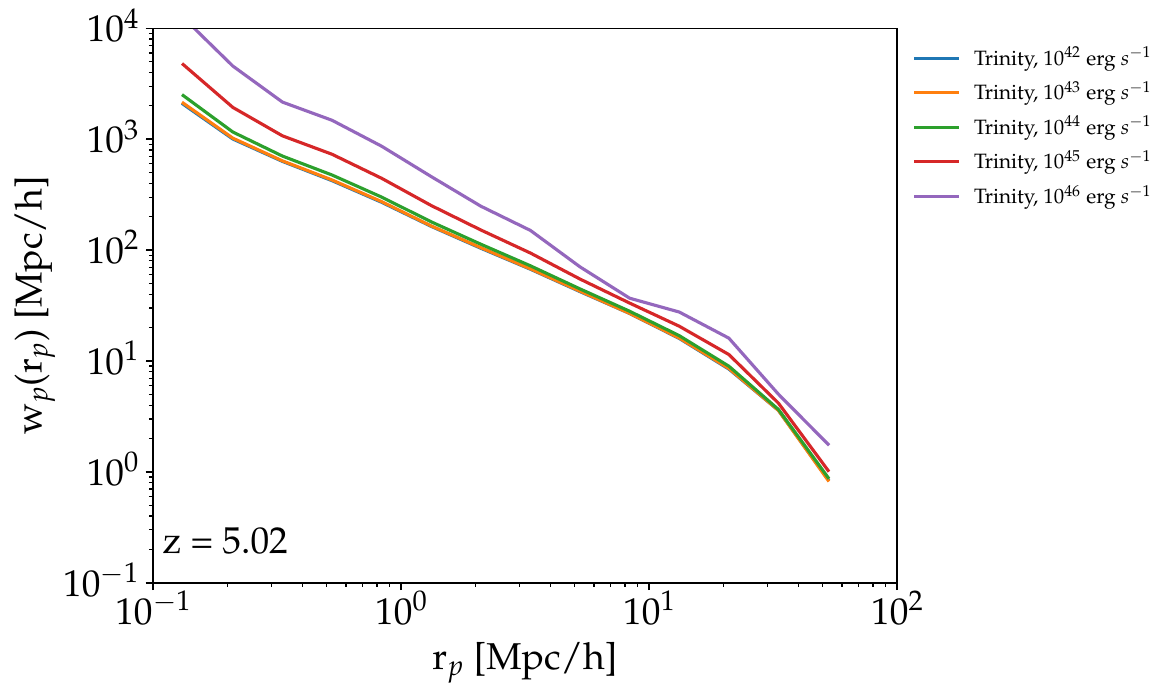} \\ 
    \includegraphics[width = \columnwidth]{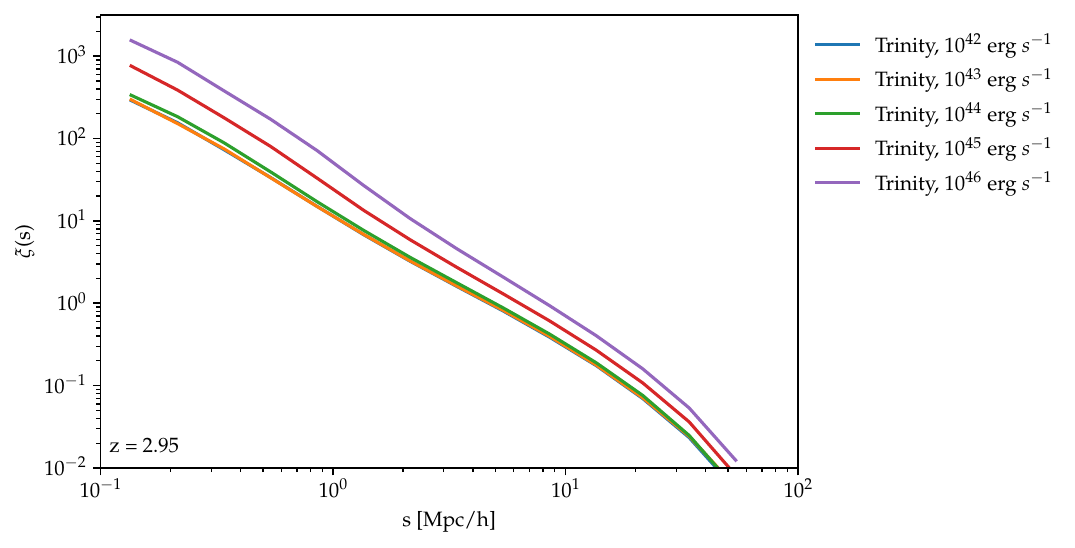} 
    \includegraphics[width = \columnwidth]{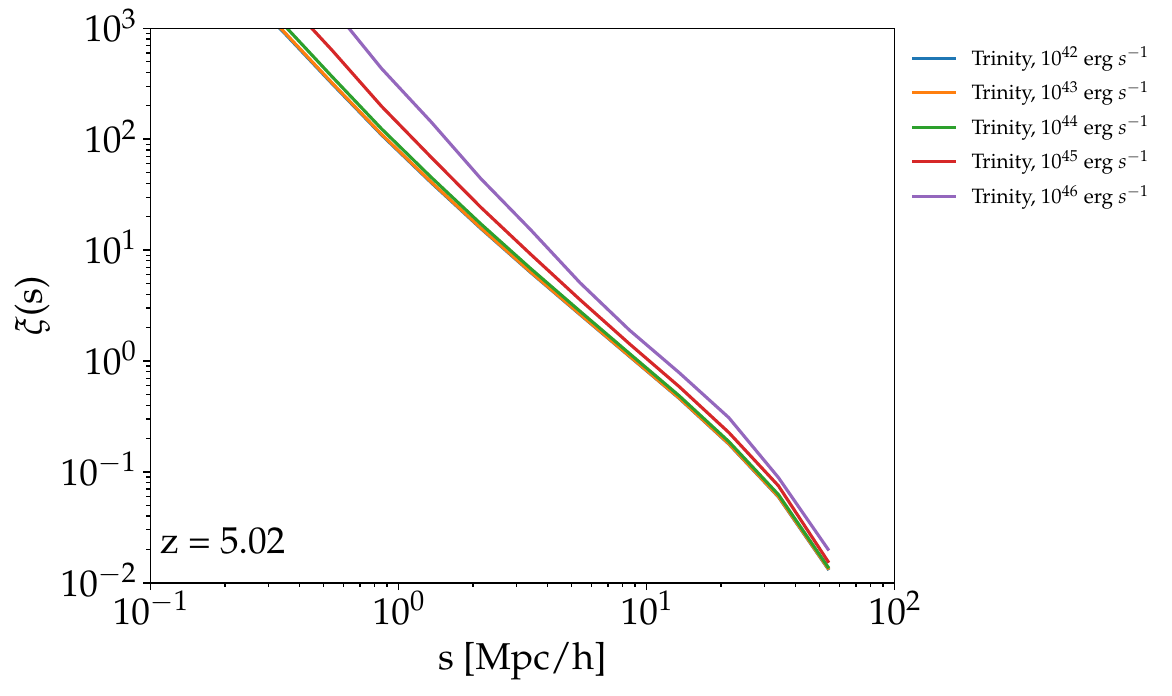}
    \caption{\textbf{Top}: Luminosity-dependent projected quasar clustering as predicted by \textsc{Trinity} at $z\sim 3$ and $z\sim 5$.   \textbf{Bottom}: Luminosity-dependent redshift-space quasar clustering as predicted by \textsc{Trinity} at $z\sim 3$ and $z\sim 5$.  In both cases, lower redshifts are shown in Appendix \ref{a:luminosities}; lower redshifts have even lower clustering differences between luminosity bins. }
    \label{fig:lum_discussion}
\end{figure*}

\label{s:discussion}

Overall, we see agreement between \textsc{Trinity} and the observed correlation functions. This is consistent with a scenario in which \textsc{Trinity} is already putting quasars in the appropriate host galaxies and host dark matter halos, without the need to use constraints from quasar clustering. We can interpret this intuitively by noting that quasar correlation functions do not change much with the luminosity threshold (see Fig.\ \ref{fig:lum_discussion} and Appendix \ref{a:luminosities}).  This arises because most of the quasars that are probed by optical surveys live in host halos and galaxies with very similar mass ranges ($M_h = 10^{12}-10^{13}\Msun$; see Paper III, \citealt{Zhang23c}, which discusses the host halos and galaxies of quasars), and hence have similar clustering. The information about host halo occupancy is most likely already present in AGN occupation fractions (i.e., the fraction of galaxies of a given mass that host AGN of a given luminosity), given the finding in \citealt{Aird2021} that AGN occupation fractions applied to galaxy mock catalogs naturally reproduce quasar biases as a function of redshift.  Indeed, AGN occupation fractions may be more constraining for physical models, as they also provide information on AGN in systems with much lower and much higher spatial clustering (and hence a wider range of host halo masses) than typical quasar samples.

Optical surveys use a wide variety of selection criteria that we cannot reproduce exactly, because there are substantial uncertainties in converting from the optical magnitude to the bolometric luminosity. Hence, different surveys could in practice be selecting quasar samples with very different bolometric luminosity distributions. Nonetheless, the observational samples with the tightest uncertainties agree both with each other and with the \textsc{Trinity} model, which is reasonable given the weak correlation between total luminosity and clustering amplitude.  

With new surveys probing fainter and fainter AGN (e.g., the BASS survey, \citealt{Powell18}, the DESI $z\gtrsim 5$ Quasar Survey, \citealt{Yang2023}, as well as upcoming \textit{Roman} and \textit{Euclid} surveys), it is helpful to ask under what conditions luminosity-dependent quasar clustering will be observed.  Fig.\ \ref{fig:lum_discussion} (with additional redshifts in Appendix \ref{a:luminosities}) provides a simple answer: \textsc{Trinity} predicts that there is typically $<0.3$ dex variation in clustering down to very low luminosities (10$^{42}$ erg/s). Most of the variation in quasar clustering occurs at high luminosities (between $10^{44-45}$ erg/s and 10$^{46}$ erg/s or greater).  But such systems are so rare that there are not enough of them to achieve tight constraints on quasar autocorrelation functions in the observable Universe.  Another message from Fig.\ \ref{fig:lum_discussion} is that the largest clustering differences are observed on small scales.  Yet, given the low number densities of quasars, accessing small scales is difficult with autocorrelation functions.  Both of these factors point to quasar-galaxy \textit{cross-correlations} as the best route to measure luminosity-dependent clustering, so that the low number density of quasars does not pose an impediment to measuring clustering at high luminosities and small scales.

We can also speculate on why quasar clustering does not seem to depend on black hole mass (e.g., \citealt{Krolewski18}).  While in part this is due to uncertain virial estimates of black hole mass (see \citealt{Krolewski18}), another factor is that selecting quasars already restricts the host halo mass range of the systems (see Paper III, \citealt{Zhang23c}).  Although we may expect lower-mass black holes to be hosted by lower-mass galaxies and halos, this is only true for the population as a whole.  The tiny fraction of black holes that are accreting rapidly enough to shine as quasars are primarily living in $10^{12}-10^{13}\Msun$ halos---and hence, further splitting the quasar population in this already narrow range of halo masses (e.g., by black hole virial mass) does not result in large clustering differences (also known as the \textit{restriction of range} effect in statistics).  

\section{Conclusions}

\label{s:conclusions}

In this paper, we compared predicted quasar autocorrelation functions from the \textsc{Trinity} model with observed quasar autocorrelation functions for $0<z<3.5$.  Key findings include:
\begin{itemize}
    \item \textsc{Trinity} agrees well with observed quasar autocorrelation functions, including their spatial dependence (Section \ref{s:results}).  This is consistent with the interpretation that \textsc{Trinity} is placing quasars in the appropriate galaxies and host halos (Section \ref{s:discussion}).
    \item As no quasar correlation functions were used to constrain \textsc{Trinity}, this implies that galaxy and host halo occupation information was already present in another observational data set used for \textsc{Trinity}--most likely AGN occupation fractions as a function of galaxy mass and redshift (Section \ref{s:discussion}).
    \item \textsc{Trinity} predicts very shallow luminosity dependence for quasar clustering (Section \ref{s:discussion}), in agreement with observations (Appendix \ref{a:luminosities}).  This is due to most quasars being hosted by halos in a very narrow mass range ($10^{12}-10^{13}\Msun$, almost regardless of selection).
    \item Future observations that target very bright quasars ($10^{46}-10^{48}$ erg/s in bolometric luminosity) may have more success in finding luminosity dependent clustering by measuring small scale ($<5$ Mpc) galaxy--quasar cross correlation functions, as quasar--quasar autocorrelation functions will be limited by low statistical power (Section \ref{s:discussion}).
\end{itemize}

\section*{Acknowledgements}

OK would like to thank Bachem and Archie Knox for their unconditional patience through this process. OK would also like to thank Ann Zabludoff for her unwavering support on OK's journey to becoming a scientist. PB was funded by a Packard Fellowship, Grant \#2019-69646.




\bibliographystyle{mnras}
\bibliography{example} 

\begin{thebibliography}{}
\makeatletter
\relax
\def\mn@urlcharsother{\let\do\@makeother \do\$\do\&\do\#\do\^\do\_\do\%\do\~}
\def\mn@doi{\begingroup\mn@urlcharsother \@ifnextchar [ {\mn@doi@} {\mn@doi@[]}}
\def\mn@doi@[#1]#2{\def\@tempa{#1}\ifx\@tempa\@empty \href {http://dx.doi.org/#2} {doi:#2}\else \href {http://dx.doi.org/#2} {#1}\fi \endgroup}
\def\mn@eprint#1#2{\mn@eprint@#1:#2::\@nil}
\def\mn@eprint@arXiv#1{\href {http://arxiv.org/abs/#1} {{\tt arXiv:#1}}}
\def\mn@eprint@dblp#1{\href {http://dblp.uni-trier.de/rec/bibtex/#1.xml} {dblp:#1}}
\def\mn@eprint@#1:#2:#3:#4\@nil{\def\@tempa {#1}\def\@tempb {#2}\def\@tempc {#3}\ifx \@tempc \@empty \let \@tempc \@tempb \let \@tempb \@tempa \fi \ifx \@tempb \@empty \def\@tempb {arXiv}\fi \@ifundefined {mn@eprint@\@tempb}{\@tempb:\@tempc}{\expandafter \expandafter \csname mn@eprint@\@tempb\endcsname \expandafter{\@tempc}}}

\bibitem[\protect\citeauthoryear{{Aird} \& {Coil}}{{Aird} \& {Coil}}{2021}]{Aird2021}
{Aird} J.,  {Coil} A.~L.,  2021, \mn@doi [\mnras] {10.1093/mnras/stab312}, \href {https://ui.adsabs.harvard.edu/abs/2021MNRAS.502.5962A} {502, 5962}

\bibitem[\protect\citeauthoryear{{Aird}, {Coil}  \& {Georgakakis}}{{Aird} et~al.}{2018}]{Aird2018}
{Aird} J.,  {Coil} A.~L.,   {Georgakakis} A.,  2018, \mn@doi [\mnras] {10.1093/mnras/stx2700}, \href {https://ui.adsabs.harvard.edu/abs/2018MNRAS.474.1225A} {474, 1225}

\bibitem[\protect\citeauthoryear{{Behroozi}, {Wechsler}  \& {Wu}}{{Behroozi} et~al.}{2013a}]{Behroozi13a}
{Behroozi} P.~S.,  {Wechsler} R.~H.,   {Wu} H.-Y.,  2013a, \mn@doi [\apj] {10.1088/0004-637X/762/2/109}, \href {https://ui.adsabs.harvard.edu/abs/2013ApJ...762..109B} {762, 109}

\bibitem[\protect\citeauthoryear{{Behroozi}, {Wechsler}, {Wu}, {Busha}, {Klypin}  \& {Primack}}{{Behroozi} et~al.}{2013b}]{Behroozi13b}
{Behroozi} P.~S.,  {Wechsler} R.~H.,  {Wu} H.-Y.,  {Busha} M.~T.,  {Klypin} A.~A.,   {Primack} J.~R.,  2013b, \mn@doi [\apj] {10.1088/0004-637X/763/1/18}, \href {https://ui.adsabs.harvard.edu/abs/2013ApJ...763...18B} {763, 18}

\bibitem[\protect\citeauthoryear{{Behroozi}, {Wechsler}, {Hearin}  \& {Conroy}}{{Behroozi} et~al.}{2019}]{Behroozi2019}
{Behroozi} P.,  {Wechsler} R.~H.,  {Hearin} A.~P.,   {Conroy} C.,  2019, \mn@doi [\mnras] {10.1093/mnras/stz1182}, \href {https://ui.adsabs.harvard.edu/abs/2019MNRAS.488.3143B} {488, 3143}

\bibitem[\protect\citeauthoryear{{Bongiorno} et~al.,}{{Bongiorno} et~al.}{2012}]{Bongiorno2012}
{Bongiorno} A.,  et~al., 2012, \mn@doi [\mnras] {10.1111/j.1365-2966.2012.22089.x}, \href {https://ui.adsabs.harvard.edu/abs/2012MNRAS.427.3103B} {427, 3103}

\bibitem[\protect\citeauthoryear{{Chehade} et~al.,}{{Chehade} et~al.}{2016}]{Chehade16}
{Chehade} B.,  et~al., 2016, \mn@doi [\mnras] {10.1093/mnras/stw616}, \href {https://ui.adsabs.harvard.edu/abs/2016MNRAS.459.1179C} {459, 1179}

\bibitem[\protect\citeauthoryear{{Conroy} \& {White}}{{Conroy} \& {White}}{2013}]{Conroy2013}
{Conroy} C.,  {White} M.,  2013, \mn@doi [\apj] {10.1088/0004-637X/762/2/70}, \href {https://ui.adsabs.harvard.edu/abs/2013ApJ...762...70C} {762, 70}

\bibitem[\protect\citeauthoryear{{Croom} et~al.,}{{Croom} et~al.}{2005}]{Croom05}
{Croom} S.~M.,  et~al., 2005, \mn@doi [\mnras] {10.1111/j.1365-2966.2004.08379.x}, \href {https://ui.adsabs.harvard.edu/abs/2005MNRAS.356..415C} {356, 415}

\bibitem[\protect\citeauthoryear{{Croton}}{{Croton}}{2009}]{Croton2009}
{Croton} D.~J.,  2009, \mn@doi [\mnras] {10.1111/j.1365-2966.2009.14429.x}, \href {https://ui.adsabs.harvard.edu/abs/2009MNRAS.394.1109C} {394, 1109}

\bibitem[\protect\citeauthoryear{{Dawson} et~al.,}{{Dawson} et~al.}{2016}]{Dawson16}
{Dawson} K.~S.,  et~al., 2016, \mn@doi [\aj] {10.3847/0004-6256/151/2/44}, \href {https://ui.adsabs.harvard.edu/abs/2016AJ....151...44D} {151, 44}

\bibitem[\protect\citeauthoryear{{Eftekharzadeh} et~al.,}{{Eftekharzadeh} et~al.}{2015}]{Eftekharzadeh15}
{Eftekharzadeh} S.,  et~al., 2015, \mn@doi [\mnras] {10.1093/mnras/stv1763}, \href {https://ui.adsabs.harvard.edu/abs/2015MNRAS.453.2779E} {453, 2779}

\bibitem[\protect\citeauthoryear{{Eisenstein} et~al.,}{{Eisenstein} et~al.}{2011}]{Eisenstein11}
{Eisenstein} D.~J.,  et~al., 2011, \mn@doi [\aj] {10.1088/0004-6256/142/3/72}, \href {https://ui.adsabs.harvard.edu/abs/2011AJ....142...72E} {142, 72}

\bibitem[\protect\citeauthoryear{{Harikane}, {Nakajima}, {Ouchi}, {Umeda}, {Isobe}, {Ono}, {Xu}  \& {Zhang}}{{Harikane} et~al.}{2023}]{Harikane2023}
{Harikane} Y.,  {Nakajima} K.,  {Ouchi} M.,  {Umeda} H.,  {Isobe} Y.,  {Ono} Y.,  {Xu} Y.,   {Zhang} Y.,  2023, \mn@doi [arXiv e-prints] {10.48550/arXiv.2304.06658}, \href {https://ui.adsabs.harvard.edu/abs/2023arXiv230406658H} {p. arXiv:2304.06658}

\bibitem[\protect\citeauthoryear{{H{\"a}ring} \& {Rix}}{{H{\"a}ring} \& {Rix}}{2004}]{Haring2004}
{H{\"a}ring} N.,  {Rix} H.-W.,  2004, \mn@doi [\apjl] {10.1086/383567}, \href {https://ui.adsabs.harvard.edu/abs/2004ApJ...604L..89H} {604, L89}

\bibitem[\protect\citeauthoryear{{Hopkins}, {Hernquist}, {Cox}  \& {Kere{\v{s}}}}{{Hopkins} et~al.}{2008}]{Hopkins2008}
{Hopkins} P.~F.,  {Hernquist} L.,  {Cox} T.~J.,   {Kere{\v{s}}} D.,  2008, \mn@doi [\apjs] {10.1086/524362}, \href {https://ui.adsabs.harvard.edu/abs/2008ApJS..175..356H} {175, 356}

\bibitem[\protect\citeauthoryear{{Klypin}, {Yepes}, {Gottl{\"o}ber}, {Prada}  \& {He{\ss}}}{{Klypin} et~al.}{2016}]{Klypin16}
{Klypin} A.,  {Yepes} G.,  {Gottl{\"o}ber} S.,  {Prada} F.,   {He{\ss}} S.,  2016, \mn@doi [\mnras] {10.1093/mnras/stw248}, \href {https://ui.adsabs.harvard.edu/abs/2016MNRAS.457.4340K} {457, 4340}

\bibitem[\protect\citeauthoryear{{Kormendy} \& {Ho}}{{Kormendy} \& {Ho}}{2013}]{Kormendy2013}
{Kormendy} J.,  {Ho} L.~C.,  2013, \mn@doi [\araa] {10.1146/annurev-astro-082708-101811}, \href {https://ui.adsabs.harvard.edu/abs/2013ARA&A..51..511K} {51, 511}

\bibitem[\protect\citeauthoryear{{Krolewski} \& {Eisenstein}}{{Krolewski} \& {Eisenstein}}{2015}]{Krolewski18}
{Krolewski} A.~G.,  {Eisenstein} D.~J.,  2015, \mn@doi [\apj] {10.1088/0004-637X/803/1/4}, \href {https://ui.adsabs.harvard.edu/abs/2015ApJ...803....4K} {803, 4}

\bibitem[\protect\citeauthoryear{{Laurent} et~al.,}{{Laurent} et~al.}{2017}]{Laurent17}
{Laurent} P.,  et~al., 2017, \mn@doi [\jcap] {10.1088/1475-7516/2017/07/017}, \href {https://ui.adsabs.harvard.edu/abs/2017JCAP...07..017L} {2017, 017}

\bibitem[\protect\citeauthoryear{{Li} et~al.,}{{Li} et~al.}{2021}]{Li21}
{Li} J.,  et~al., 2021, \mn@doi [\apj] {10.3847/1538-4357/ac2301}, \href {https://ui.adsabs.harvard.edu/abs/2021ApJ...922..142L} {922, 142}

\bibitem[\protect\citeauthoryear{{McConnell} \& {Ma}}{{McConnell} \& {Ma}}{2013}]{McConnell2013}
{McConnell} N.~J.,  {Ma} C.-P.,  2013, \mn@doi [\apj] {10.1088/0004-637X/764/2/184}, \href {https://ui.adsabs.harvard.edu/abs/2013ApJ...764..184M} {764, 184}

\bibitem[\protect\citeauthoryear{{Moster}, {Naab}  \& {White}}{{Moster} et~al.}{2018}]{Moster18}
{Moster} B.~P.,  {Naab} T.,   {White} S. D.~M.,  2018, \mn@doi [\mnras] {10.1093/mnras/sty655}, \href {https://ui.adsabs.harvard.edu/abs/2018MNRAS.477.1822M} {477, 1822}

\bibitem[\protect\citeauthoryear{{Naab} \& {Ostriker}}{{Naab} \& {Ostriker}}{2017}]{Naab17}
{Naab} T.,  {Ostriker} J.~P.,  2017, \mn@doi [\araa] {10.1146/annurev-astro-081913-040019}, \href {https://ui.adsabs.harvard.edu/abs/2017ARA&A..55...59N} {55, 59}

\bibitem[\protect\citeauthoryear{{Planck Collaboration} et~al.,}{{Planck Collaboration} et~al.}{2016}]{Planck2016}
{Planck Collaboration} et~al., 2016, \mn@doi [\aap] {10.1051/0004-6361/201525830}, \href {https://ui.adsabs.harvard.edu/abs/2016A&A...594A..13P} {594, A13}

\bibitem[\protect\citeauthoryear{{Powell} et~al.,}{{Powell} et~al.}{2018}]{Powell18}
{Powell} M.~C.,  et~al., 2018, \mn@doi [\apj] {10.3847/1538-4357/aabd7f}, \href {https://ui.adsabs.harvard.edu/abs/2018ApJ...858..110P} {858, 110}

\bibitem[\protect\citeauthoryear{{Powell}, {Krumpe}, {Coil}  \& {Miyaji}}{{Powell} et~al.}{2024}]{Powell2024}
{Powell} M.~C.,  {Krumpe} M.,  {Coil} A.,   {Miyaji} T.,  2024, \mn@doi [arXiv e-prints] {10.48550/arXiv.2403.03978}, \href {https://ui.adsabs.harvard.edu/abs/2024arXiv240303978P} {p. arXiv:2403.03978}

\bibitem[\protect\citeauthoryear{{Richards} et~al.,}{{Richards} et~al.}{2005}]{Richards05}
{Richards} G.~T.,  et~al., 2005, \mn@doi [\mnras] {10.1111/j.1365-2966.2005.09096.x}, \href {https://ui.adsabs.harvard.edu/abs/2005MNRAS.360..839R} {360, 839}

\bibitem[\protect\citeauthoryear{{Rodr{\'\i}guez-Puebla}, {Behroozi}, {Primack}, {Klypin}, {Lee}  \& {Hellinger}}{{Rodr{\'\i}guez-Puebla} et~al.}{2016}]{Rodriguez-Puebla16}
{Rodr{\'\i}guez-Puebla} A.,  {Behroozi} P.,  {Primack} J.,  {Klypin} A.,  {Lee} C.,   {Hellinger} D.,  2016, \mn@doi [\mnras] {10.1093/mnras/stw1705}, \href {https://ui.adsabs.harvard.edu/abs/2016MNRAS.462..893R} {462, 893}

\bibitem[\protect\citeauthoryear{{Ross} et~al.,}{{Ross} et~al.}{2009}]{Ross09}
{Ross} N.~P.,  et~al., 2009, \mn@doi [\apj] {10.1088/0004-637X/697/2/1634}, \href {https://ui.adsabs.harvard.edu/abs/2009ApJ...697.1634R} {697, 1634}

\bibitem[\protect\citeauthoryear{{Schneider} et~al.,}{{Schneider} et~al.}{2007}]{Schneider07}
{Schneider} D.~P.,  et~al., 2007, \mn@doi [\aj] {10.1086/518474}, \href {https://ui.adsabs.harvard.edu/abs/2007AJ....134..102S} {134, 102}

\bibitem[\protect\citeauthoryear{{Shankar} et~al.,}{{Shankar} et~al.}{2020a}]{Shankar2020Nat}
{Shankar} F.,  et~al., 2020a, \mn@doi [Nature Astronomy] {10.1038/s41550-019-0949-y}, \href {https://ui.adsabs.harvard.edu/abs/2020NatAs...4..282S} {4, 282}

\bibitem[\protect\citeauthoryear{{Shankar} et~al.,}{{Shankar} et~al.}{2020b}]{Shankar2020MNRAS}
{Shankar} F.,  et~al., 2020b, \mn@doi [\mnras] {10.1093/mnras/stz3522}, \href {https://ui.adsabs.harvard.edu/abs/2020MNRAS.493.1500S} {493, 1500}

\bibitem[\protect\citeauthoryear{{Shanks} et~al.,}{{Shanks} et~al.}{2015}]{Shanks15}
{Shanks} T.,  et~al., 2015, \mn@doi [\mnras] {10.1093/mnras/stv1130}, \href {https://ui.adsabs.harvard.edu/abs/2015MNRAS.451.4238S} {451, 4238}

\bibitem[\protect\citeauthoryear{{Shen}}{{Shen}}{2009}]{Shen2009model}
{Shen} Y.,  2009, \mn@doi [\apj] {10.1088/0004-637X/704/1/89}, \href {https://ui.adsabs.harvard.edu/abs/2009ApJ...704...89S} {704, 89}

\bibitem[\protect\citeauthoryear{{Shen} et~al.,}{{Shen} et~al.}{2009}]{Shen2009data}
{Shen} Y.,  et~al., 2009, \mn@doi [\apj] {10.1088/0004-637X/697/2/1656}, \href {https://ui.adsabs.harvard.edu/abs/2009ApJ...697.1656S} {697, 1656}

\bibitem[\protect\citeauthoryear{{Silk} \& {Mamon}}{{Silk} \& {Mamon}}{2012}]{Silk12}
{Silk} J.,  {Mamon} G.~A.,  2012, \mn@doi [Research in Astronomy and Astrophysics] {10.1088/1674-4527/12/8/004}, \href {https://ui.adsabs.harvard.edu/abs/2012RAA....12..917S} {12, 917}

\bibitem[\protect\citeauthoryear{{Smith}, {Croom}, {Boyle}, {Shanks}, {Miller}  \& {Loaring}}{{Smith} et~al.}{2005}]{Smith05}
{Smith} R.~J.,  {Croom} S.~M.,  {Boyle} B.~J.,  {Shanks} T.,  {Miller} L.,   {Loaring} N.~S.,  2005, \mn@doi [\mnras] {10.1111/j.1365-2966.2005.08870.x}, \href {https://ui.adsabs.harvard.edu/abs/2005MNRAS.359...57S} {359, 57}

\bibitem[\protect\citeauthoryear{{Somerville} \& {Dav{\'e}}}{{Somerville} \& {Dav{\'e}}}{2015}]{Somerville15}
{Somerville} R.~S.,  {Dav{\'e}} R.,  2015, \mn@doi [\araa] {10.1146/annurev-astro-082812-140951}, \href {https://ui.adsabs.harvard.edu/abs/2015ARA&A..53...51S} {53, 51}

\bibitem[\protect\citeauthoryear{{Tinker}, {Robertson}, {Kravtsov}, {Klypin}, {Warren}, {Yepes}  \& {Gottl{\"o}ber}}{{Tinker} et~al.}{2010}]{Tinker10}
{Tinker} J.~L.,  {Robertson} B.~E.,  {Kravtsov} A.~V.,  {Klypin} A.,  {Warren} M.~S.,  {Yepes} G.,   {Gottl{\"o}ber} S.,  2010, \mn@doi [\apj] {10.1088/0004-637X/724/2/878}, \href {https://ui.adsabs.harvard.edu/abs/2010ApJ...724..878T} {724, 878}

\bibitem[\protect\citeauthoryear{{Vogelsberger}, {Marinacci}, {Torrey}  \& {Puchwein}}{{Vogelsberger} et~al.}{2020}]{Vogelsberger20}
{Vogelsberger} M.,  {Marinacci} F.,  {Torrey} P.,   {Puchwein} E.,  2020, \mn@doi [Nature Reviews Physics] {10.1038/s42254-019-0127-2}, \href {https://ui.adsabs.harvard.edu/abs/2020NatRP...2...42V} {2, 42}

\bibitem[\protect\citeauthoryear{{Wechsler} \& {Tinker}}{{Wechsler} \& {Tinker}}{2018}]{Wechsler2018}
{Wechsler} R.~H.,  {Tinker} J.~L.,  2018, \mn@doi [\araa] {10.1146/annurev-astro-081817-051756}, \href {https://ui.adsabs.harvard.edu/abs/2018ARA&A..56..435W} {56, 435}

\bibitem[\protect\citeauthoryear{{Yang} et~al.,}{{Yang} et~al.}{2023}]{Yang2023}
{Yang} J.,  et~al., 2023, \mn@doi [\apjs] {10.3847/1538-4365/acf99b}, \href {https://ui.adsabs.harvard.edu/abs/2023ApJS..269...27Y} {269, 27}

\bibitem[\protect\citeauthoryear{{York} et~al.,}{{York} et~al.}{2000}]{York00}
{York} D.~G.,  et~al., 2000, \mn@doi [\aj] {10.1086/301513}, \href {https://ui.adsabs.harvard.edu/abs/2000AJ....120.1579Y} {120, 1579}

\bibitem[\protect\citeauthoryear{{Zhang}, {Behroozi}, {Volonteri}, {Silk}, {Fan}, {Aird}, {Yang}  \& {Hopkins}}{{Zhang} et~al.}{2023a}]{Zhang23c}
{Zhang} H.,  {Behroozi} P.,  {Volonteri} M.,  {Silk} J.,  {Fan} X.,  {Aird} J.,  {Yang} J.,   {Hopkins} P.~F.,  2023a, \mn@doi [arXiv e-prints] {10.48550/arXiv.2305.19315}, \href {https://ui.adsabs.harvard.edu/abs/2023arXiv230519315Z} {p. arXiv:2305.19315}

\bibitem[\protect\citeauthoryear{{Zhang}, {Behroozi}, {Volonteri}, {Silk}, {Fan}, {Hopkins}, {Yang}  \& {Aird}}{{Zhang} et~al.}{2023b}]{Zhang2023}
{Zhang} H.,  {Behroozi} P.,  {Volonteri} M.,  {Silk} J.,  {Fan} X.,  {Hopkins} P.~F.,  {Yang} J.,   {Aird} J.,  2023b, \mn@doi [\mnras] {10.1093/mnras/stac2633}, \href {https://ui.adsabs.harvard.edu/abs/2023MNRAS.518.2123Z} {518, 2123}

\bibitem[\protect\citeauthoryear{{da {\^A}ngela} et~al.,}{{da {\^A}ngela} et~al.}{2008}]{daAngela08}
{da {\^A}ngela} J.,  et~al., 2008, \mn@doi [\mnras] {10.1111/j.1365-2966.2007.12552.x}, 383, 565

\makeatother
\end{thebibliography}




\appendix

\section{Luminosity-Binned Quasar Clustering}
\label{a:luminosities}

\begin{figure*}
	\begin{tabular}{ll}
	\includegraphics[width = 0.5\columnwidth]{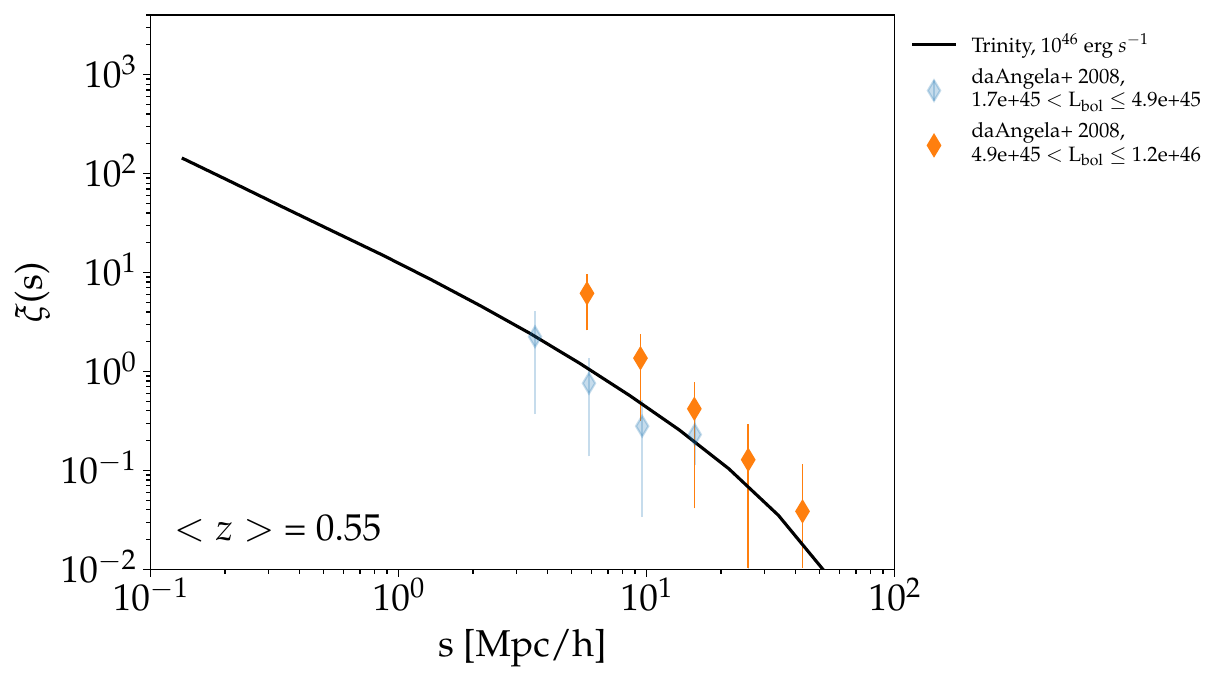} & 
    \includegraphics[width = 0.5\columnwidth]{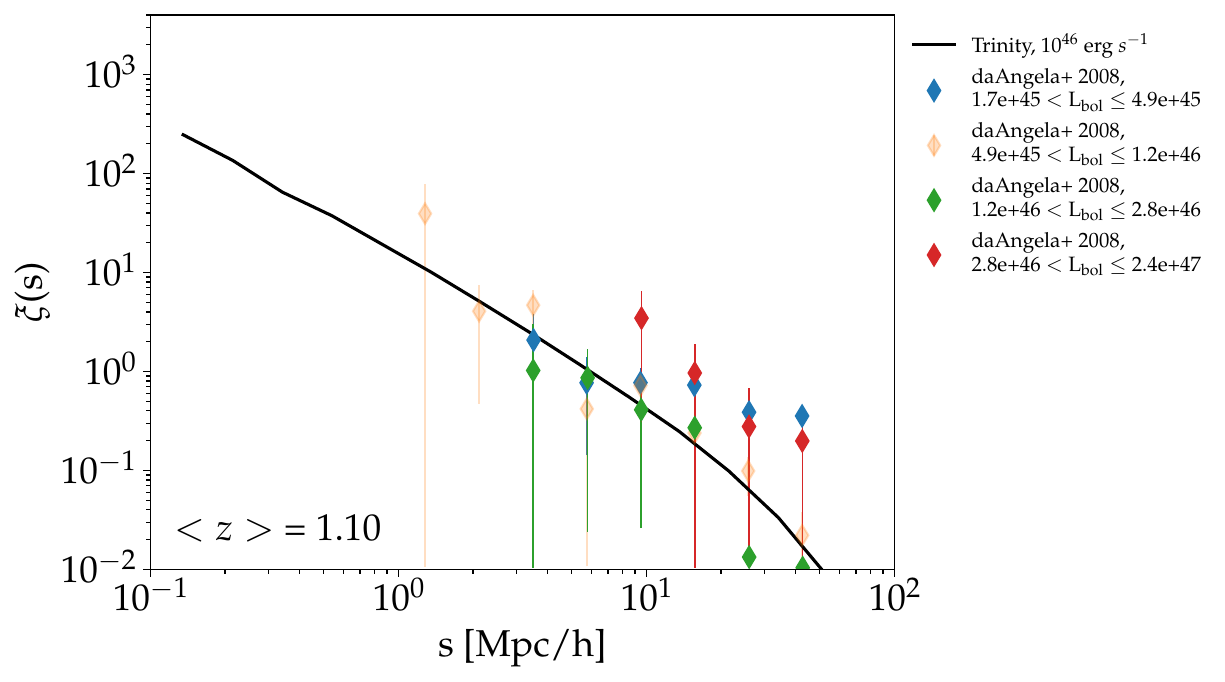}\\
    \includegraphics[width = 0.5\columnwidth]{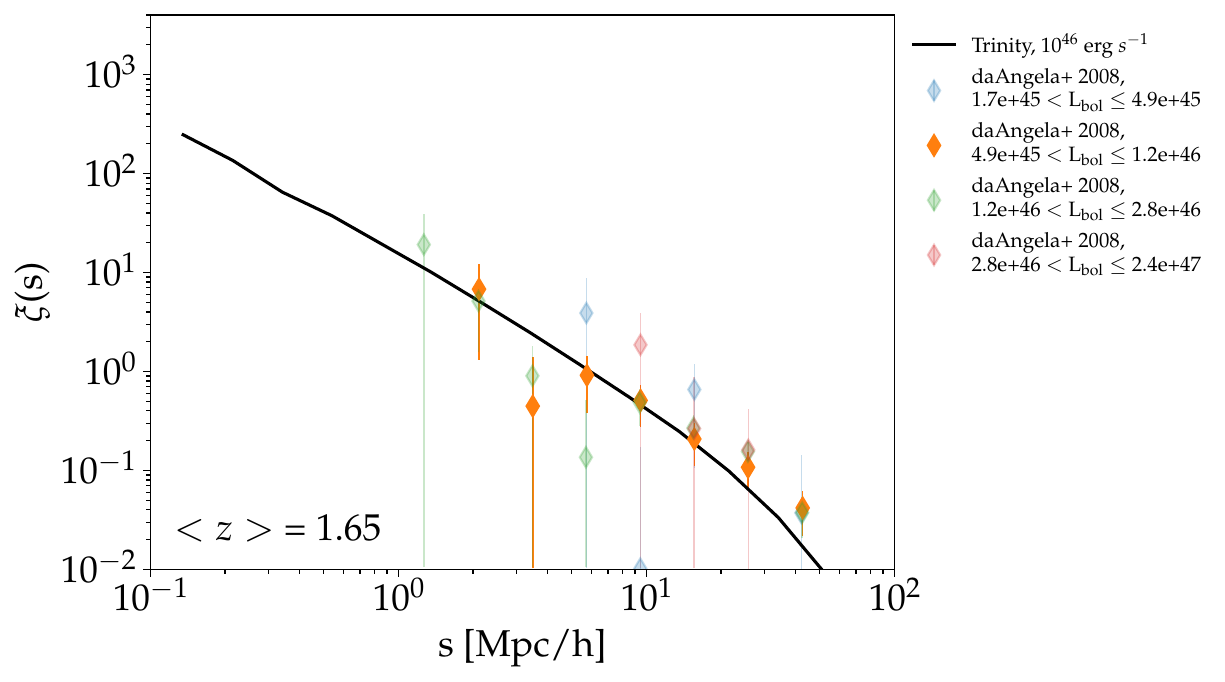} & 
    \includegraphics[width = 0.5\columnwidth]{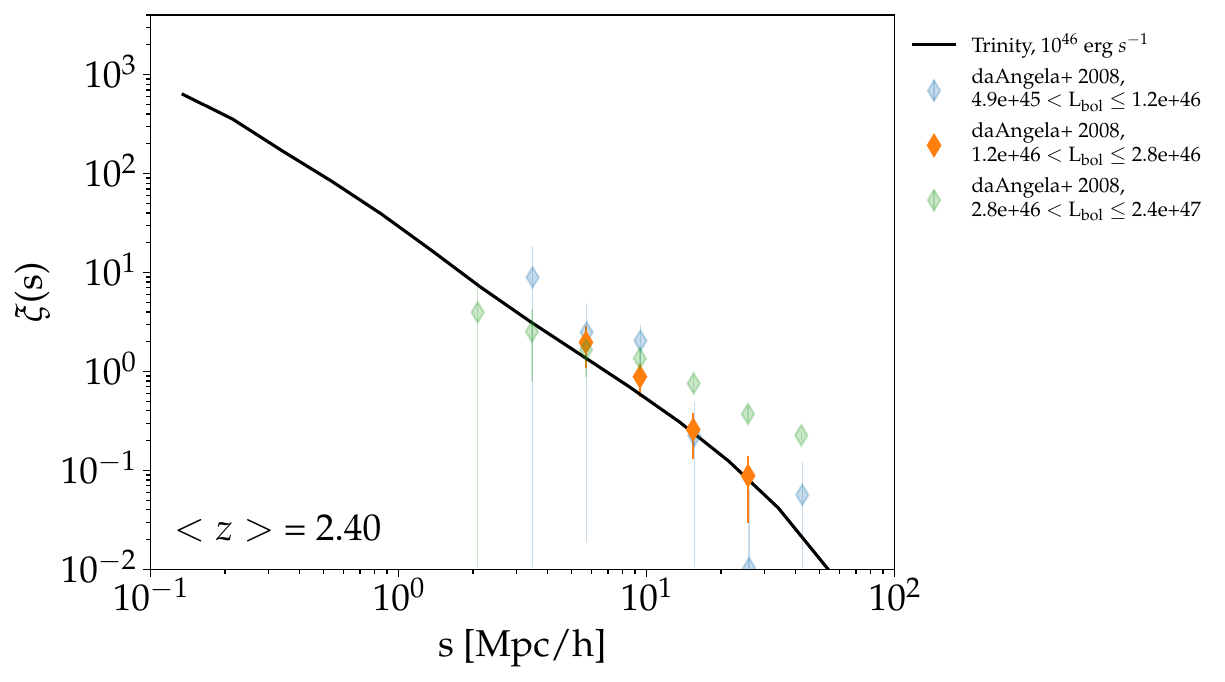}
    \end{tabular}
    \caption{The 3D redshift-space correlation function, $\xi_s$ plotted with \citet{daAngela08} luminosity-dependent observations.Observations with $>2$ dex uncertainties are shown in transparent hues so that more precise measurements stand out visually. } 
    \label{fig:obs_lum_3da}
\end{figure*}

\begin{figure*}
	\begin{tabular}{ll}
	\includegraphics[width = 0.5\columnwidth]{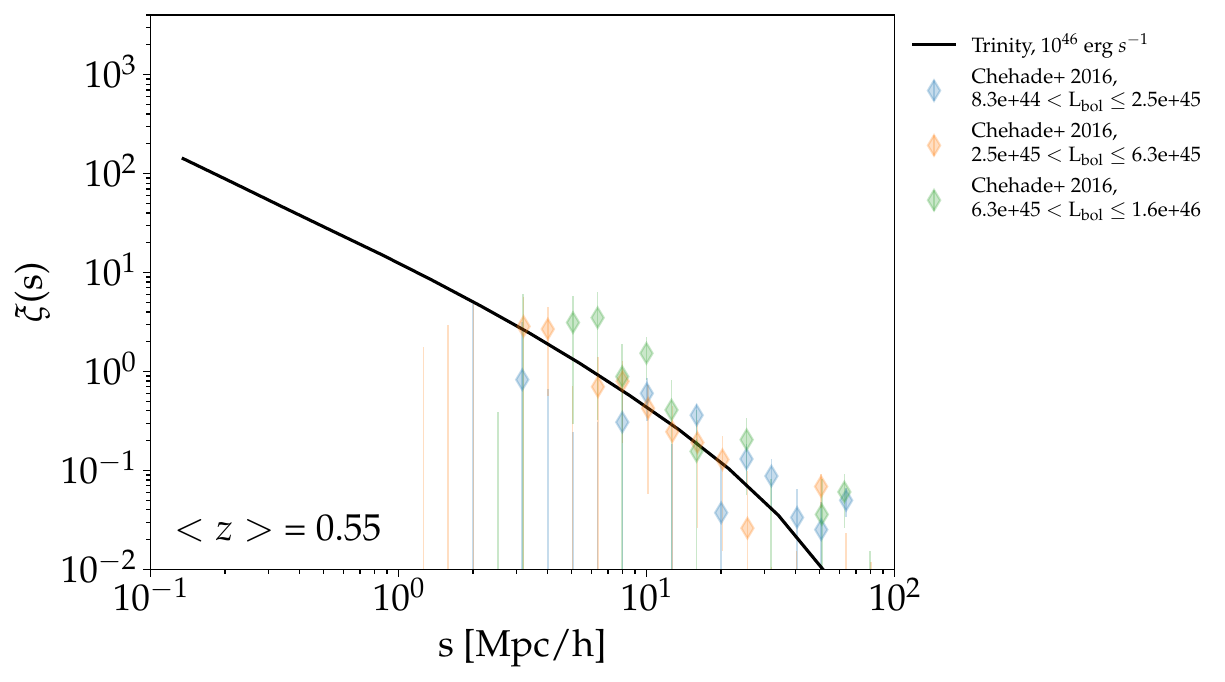} & 
	\includegraphics[width = 0.5\columnwidth]{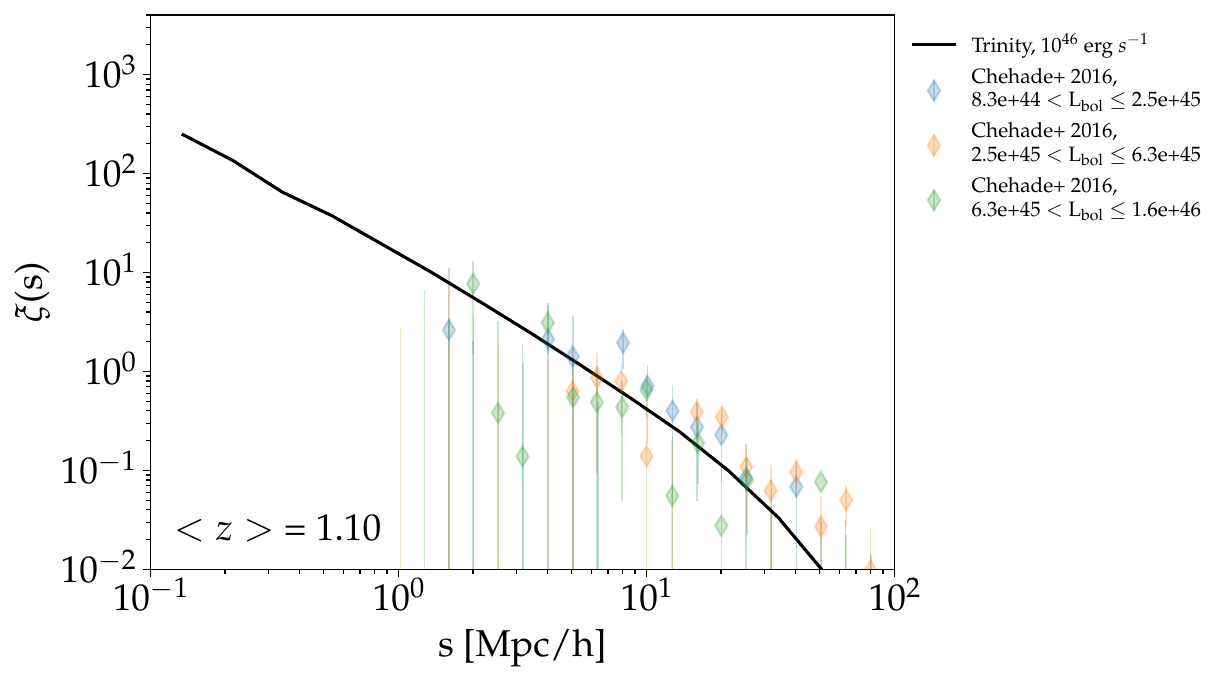}\\
	\includegraphics[width = 0.5\columnwidth]{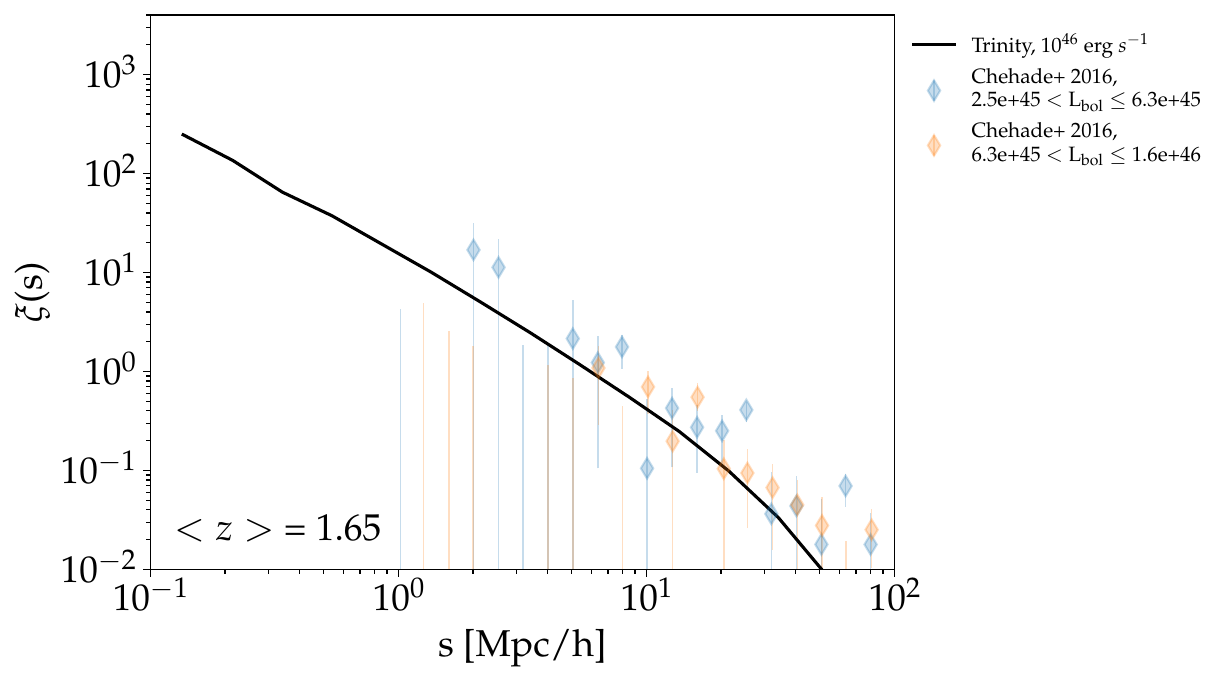} & 
	\includegraphics[width = 0.5\columnwidth]{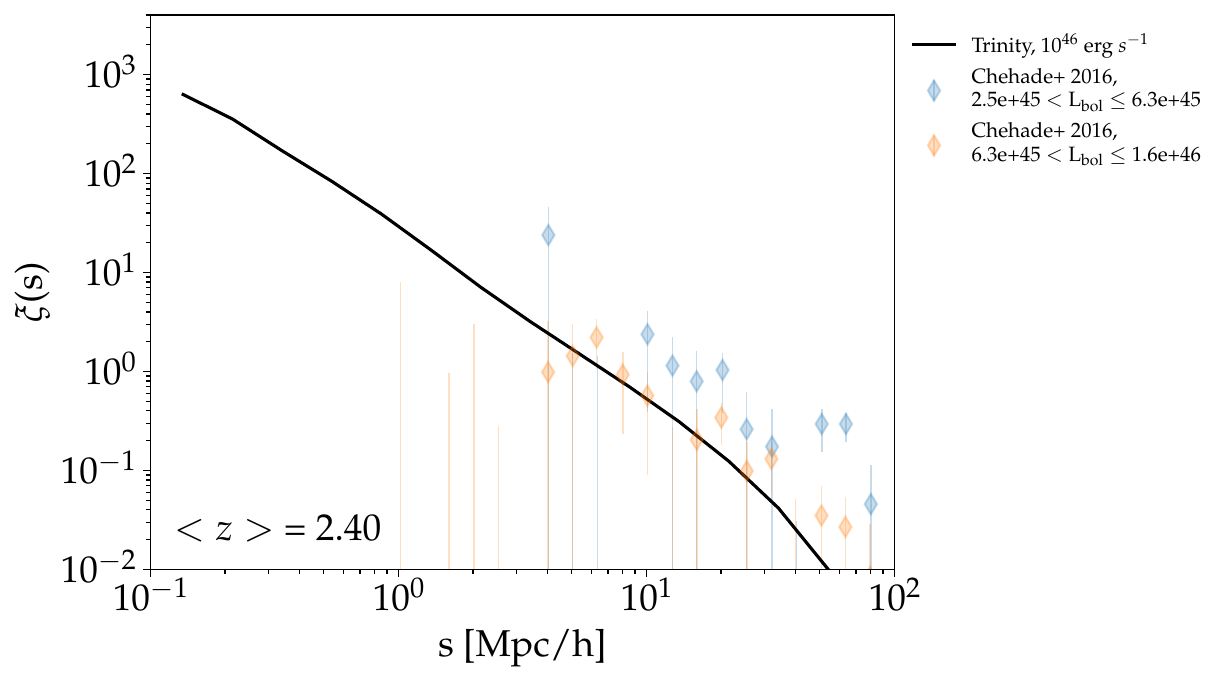}\\
	\end{tabular}
    \caption{The 3D redshift-space correlation function, $\xi_s$ plotted with \citet{Chehade16} luminosity-dependent observations. Observations with $>2$ dex uncertainties are shown in transparent hues so that more precise measurements stand out visually.}
    \label{fig:obs_lum_3db}
\end{figure*}

\begin{figure*}
    \vspace{-4cm}
    \begin{tabular}{ll}
    \includegraphics[width = 0.5\columnwidth]{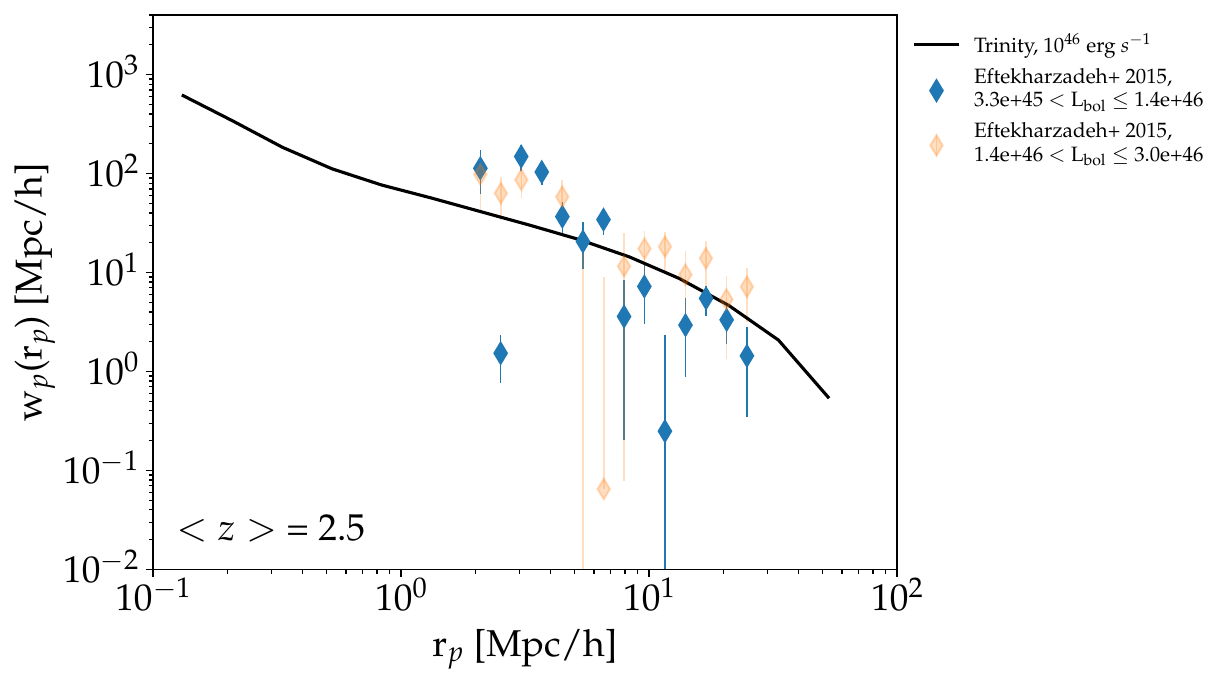} & 
    \includegraphics[width = 0.5\columnwidth]{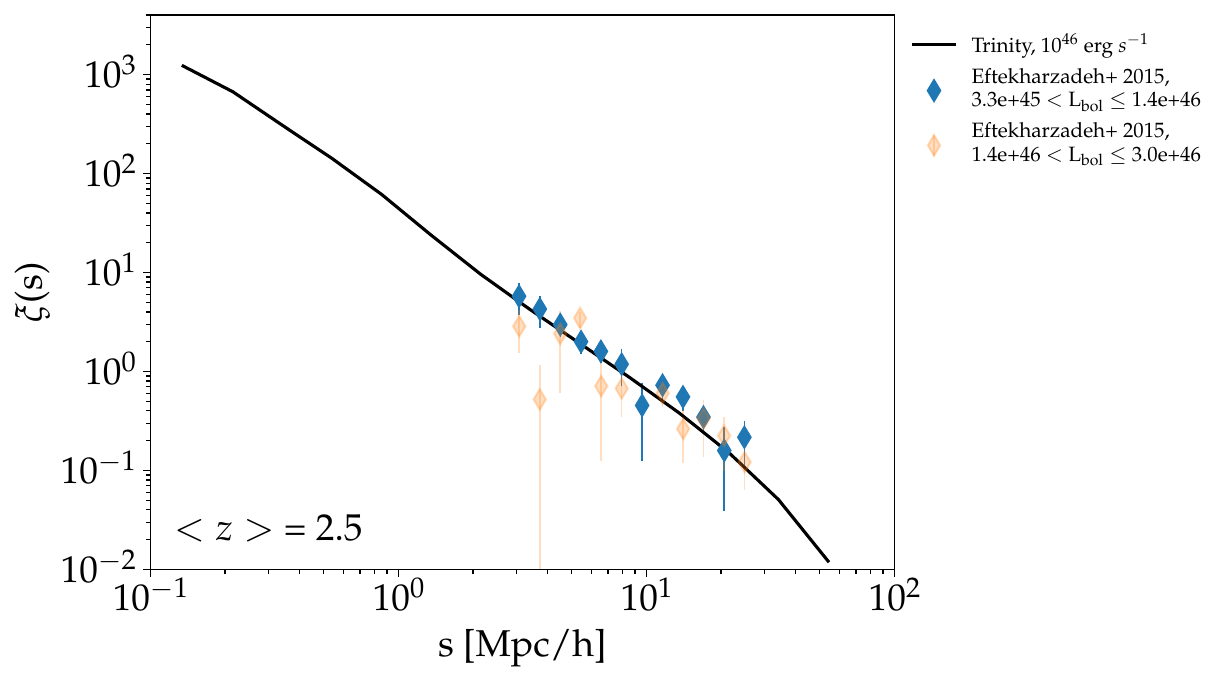} 
    \end{tabular}
    \caption{The two-point projected, $w_p$, and 3D redshift-space correlation functions, $\xi_s$ plotted with \citet{Eftekharzadeh15} luminosity-dependent observations at $<z=2.5>$. Observations with $>2$ dex uncertainties are shown in transparent hues so that more precise measurements stand out visually.}
    
    \label{fig:obs_lum_2d}
\end{figure*}

We show the \cite{daAngela08}, \cite{Chehade16}, and \cite{Eftekharzadeh15} luminosity clustering compared with \textsc{TRINITY's} predictions in Figs.\ \ref{fig:obs_lum_3da}--\ref{fig:obs_lum_2d}.  The observations show no evident luminosity dependence. The \cite{daAngela08} and \cite{Chehade16} observations of the 3D redshift-space correlation function had average redshifts of $\langle z \rangle$ = 0.55, 1.10, 1.65, 2.40, and the \cite{Eftekharzadeh15} data, for both the two-point projected and 3D redshift-space correlation function, were measured at $\langle z \rangle$ = 2.5.  As expected from splitting observed samples, the observational error bars in the luminosity-split populations are larger in all cases compared to the whole-population constraints in Figs.\ \ref{fig:2d}--\ref{fig:3d_2}, so it is not surprising that \textsc{Trinity}'s predictions continue to agree with the observations.

\begin{figure*}
	\begin{tabular}{ll}
	\includegraphics[width = 0.5\columnwidth]{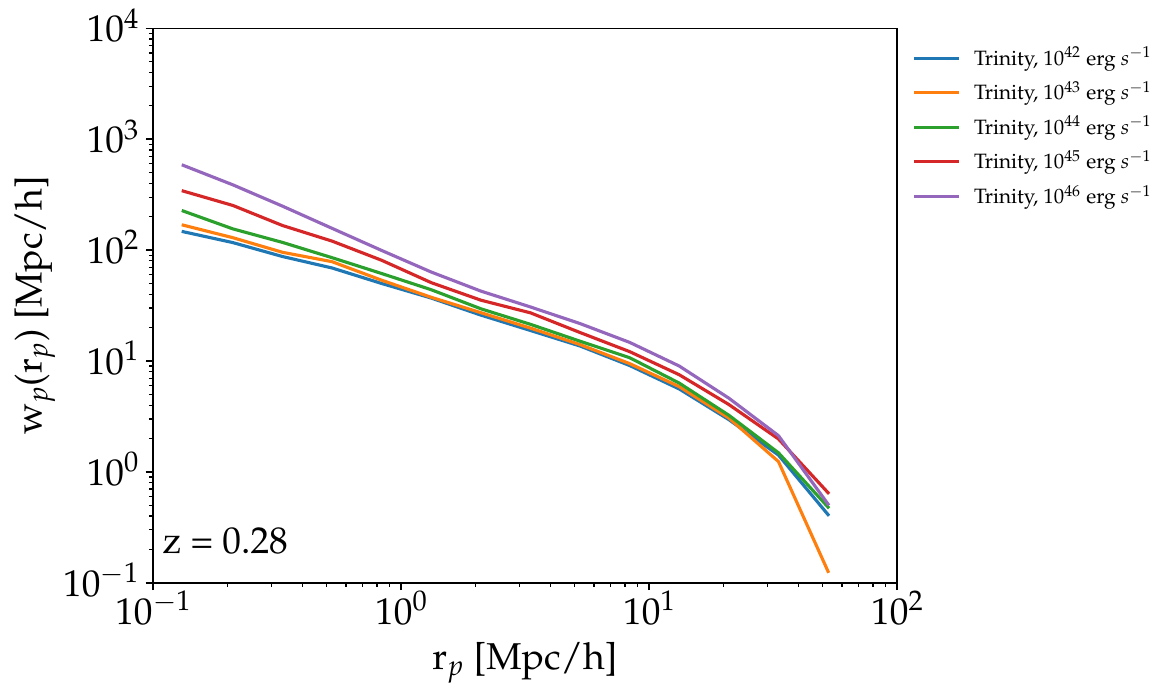} & 
	\includegraphics[width = 0.5\columnwidth]{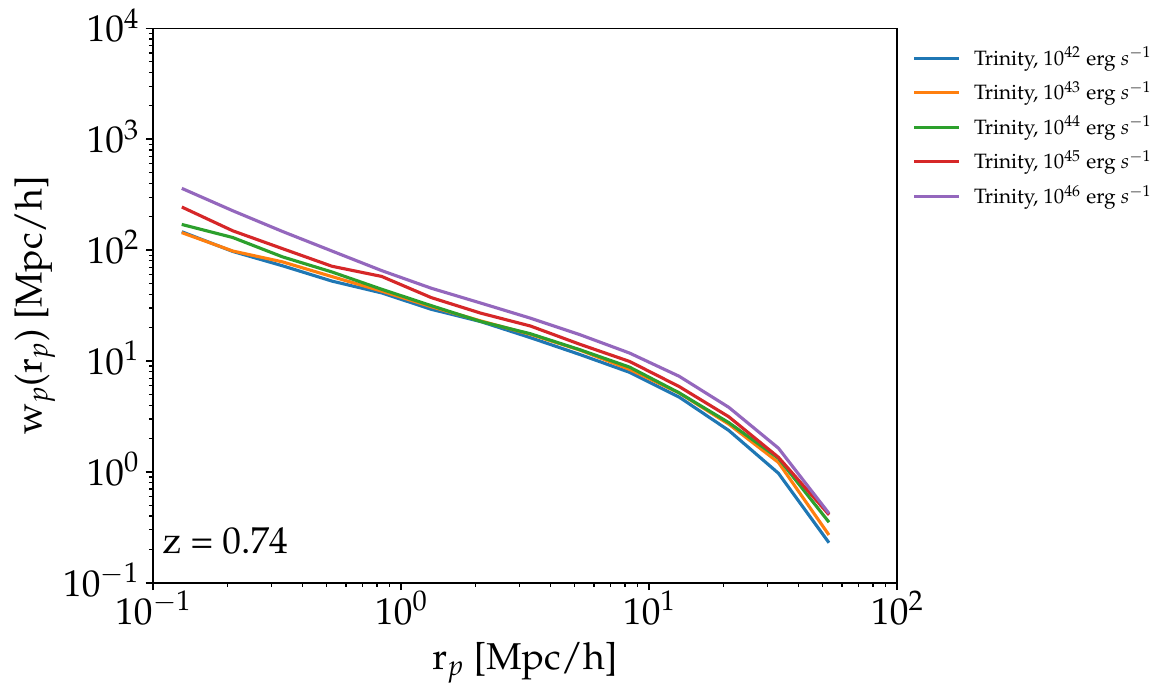}\\
	\includegraphics[width = 0.5\columnwidth]{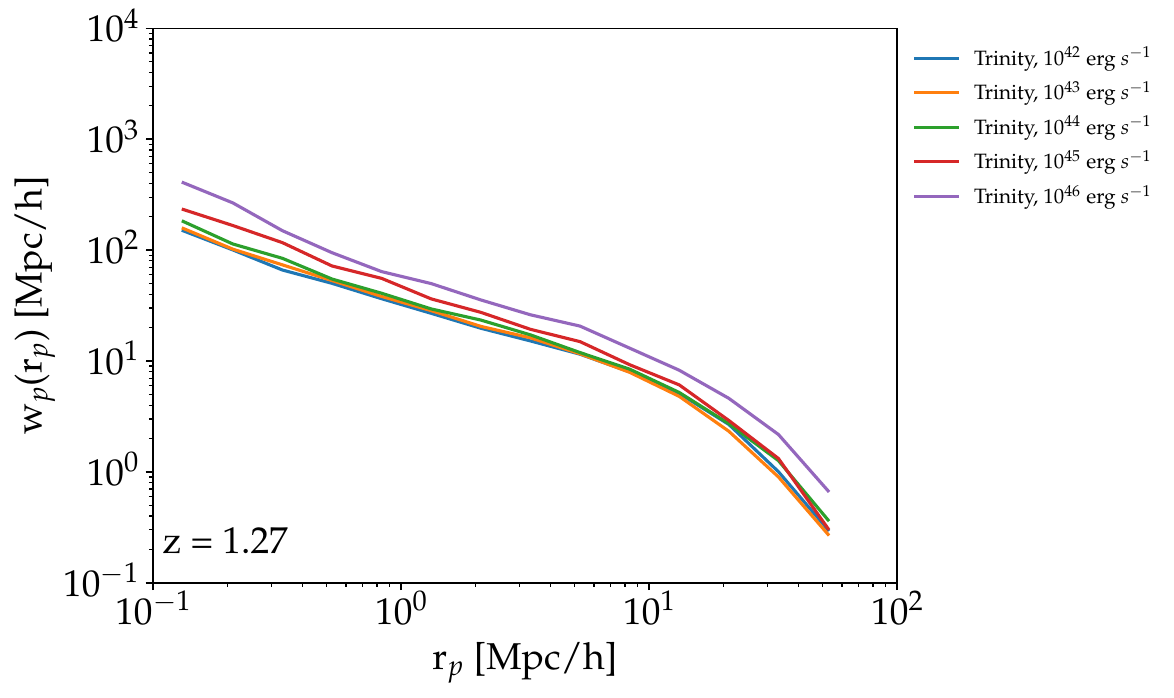} & 
	\includegraphics[width = 0.5\columnwidth]{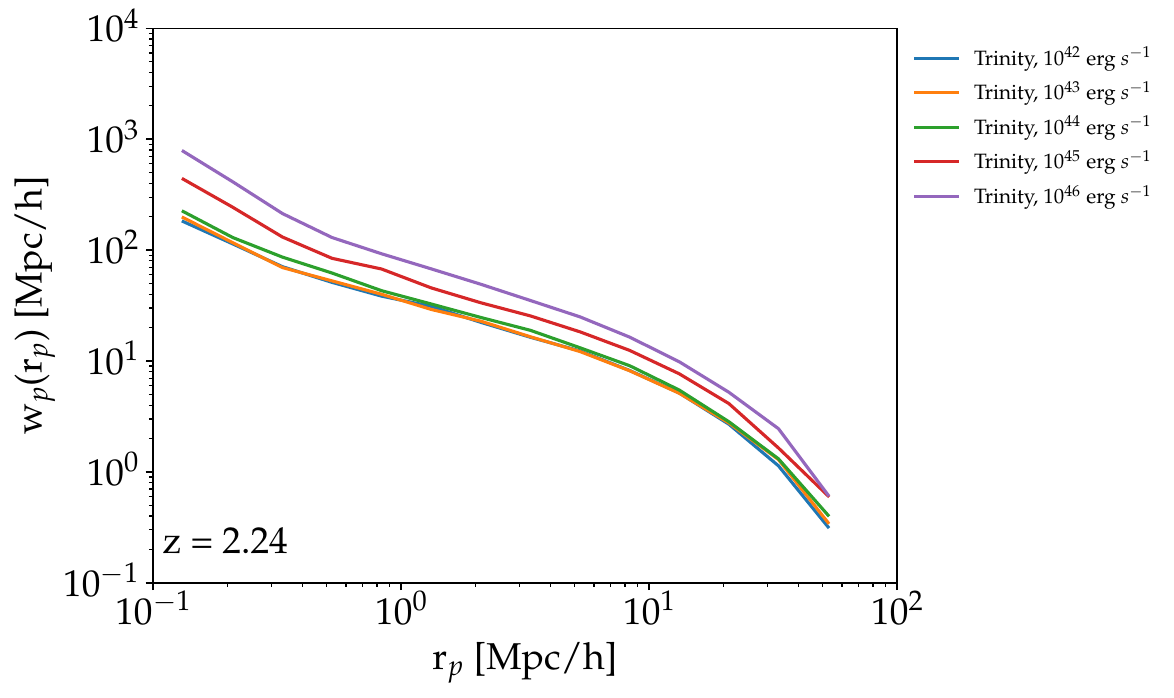}
	\end{tabular}
    \caption{The two-point projected correlation function, $w_p$, for \textsc{TRINITY} from $10^{42}$ - $10^{46}$ erg s$^{-1}$. Each luminosity bin is represented by a different color.} 
    \label{fig:lum_dependent_wp}
\end{figure*}

We show \textsc{Trinity}'s predictions for luminosity-dependent clustering at $z<2.5$ in Figs.\ \ref{fig:lum_dependent_wp} and \ref{fig:lum_dependent_xi_s}.  Compared to Fig.\ \ref{fig:lum_discussion}, lower redshifts show even less predicted luminosity dependence for quasar clustering, arising because the quasar population at lower redshifts is hosted by an even narrower typical range of halo masses than at higher redshifts \citep{Zhang23c}. For this reason, we do not show separate \textsc{Trinity} predictions for each luminosity bin in Figs.\ \ref{fig:obs_lum_3da}--\ref{fig:obs_lum_2d}.

\begin{figure*}
    \begin{tabular}{ll}
    \includegraphics[width = 0.5\columnwidth]{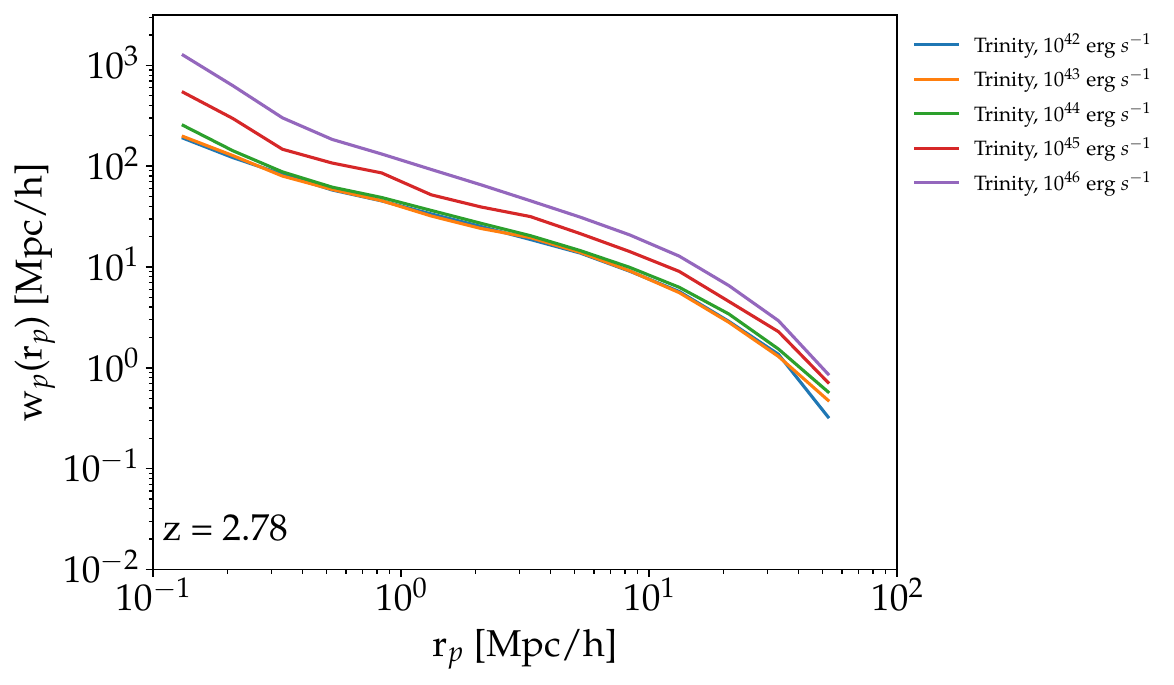} & 
    \includegraphics[width = 0.5\columnwidth]{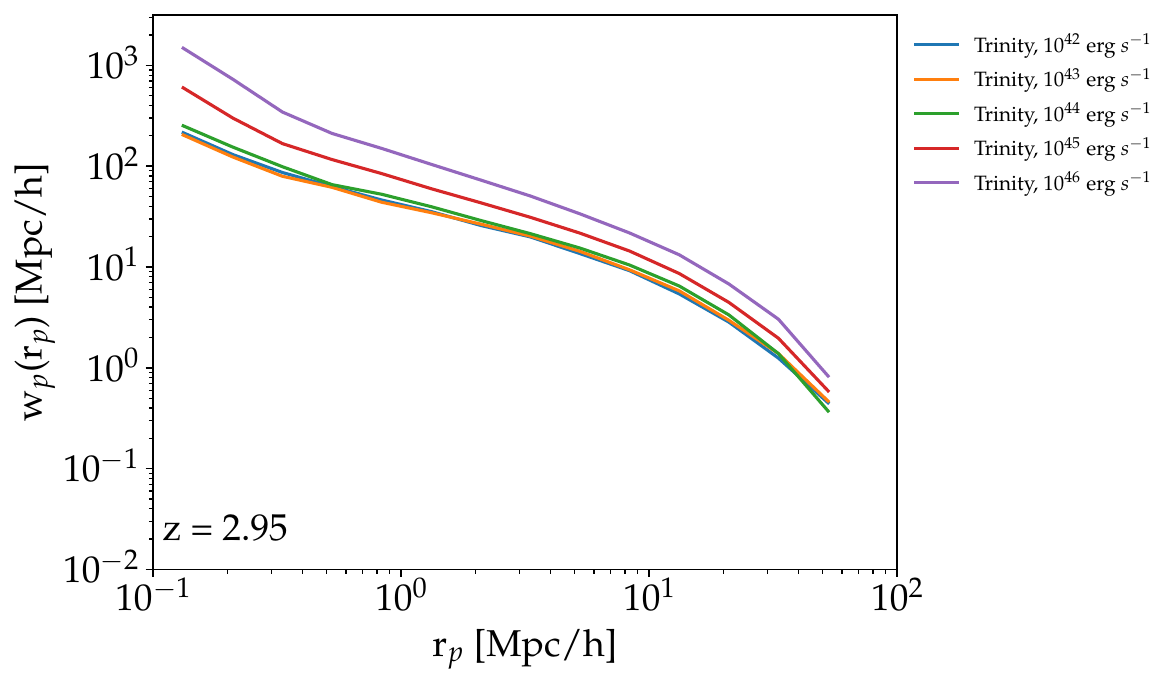}
    \end{tabular}
    \caption{This is an example figure. Captions appear below each figure.
	Give enough detail for the reader to understand what they're looking at,
	but leave detailed discussion to the main body of the text.}
    
    \label{fig:example_figure}
\end{figure*}


\begin{figure*}
	\begin{tabular}{ll}
	\includegraphics[width = 0.5\columnwidth]{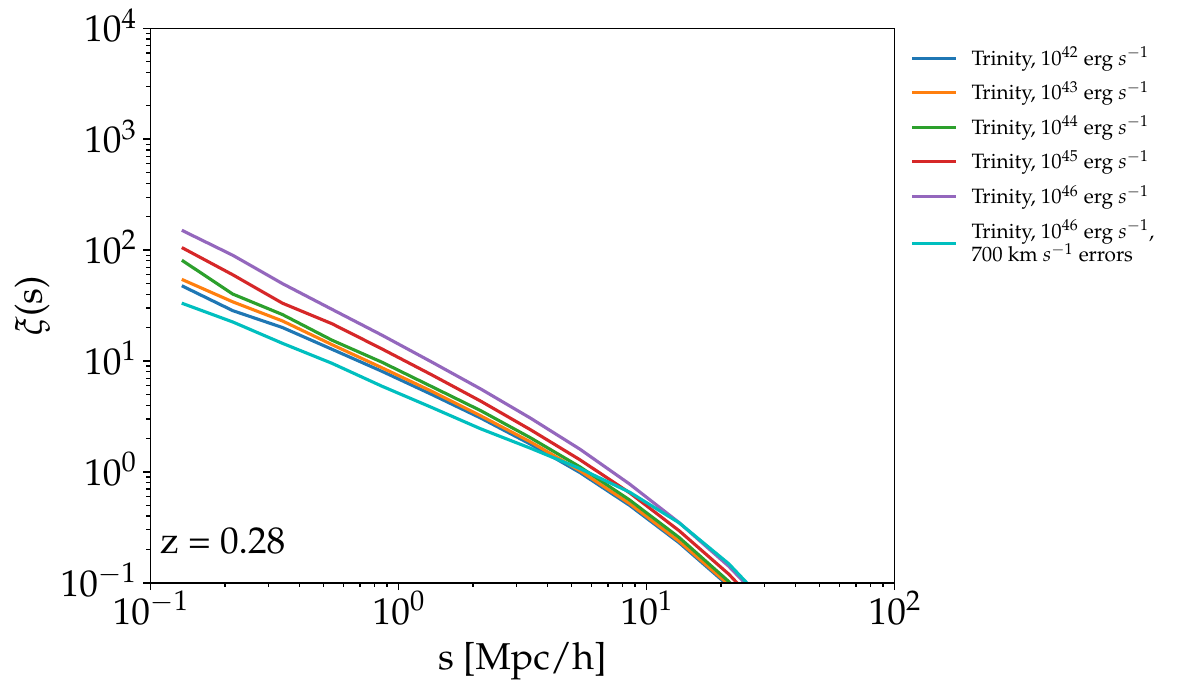} & 
	\includegraphics[width = 0.5\columnwidth]{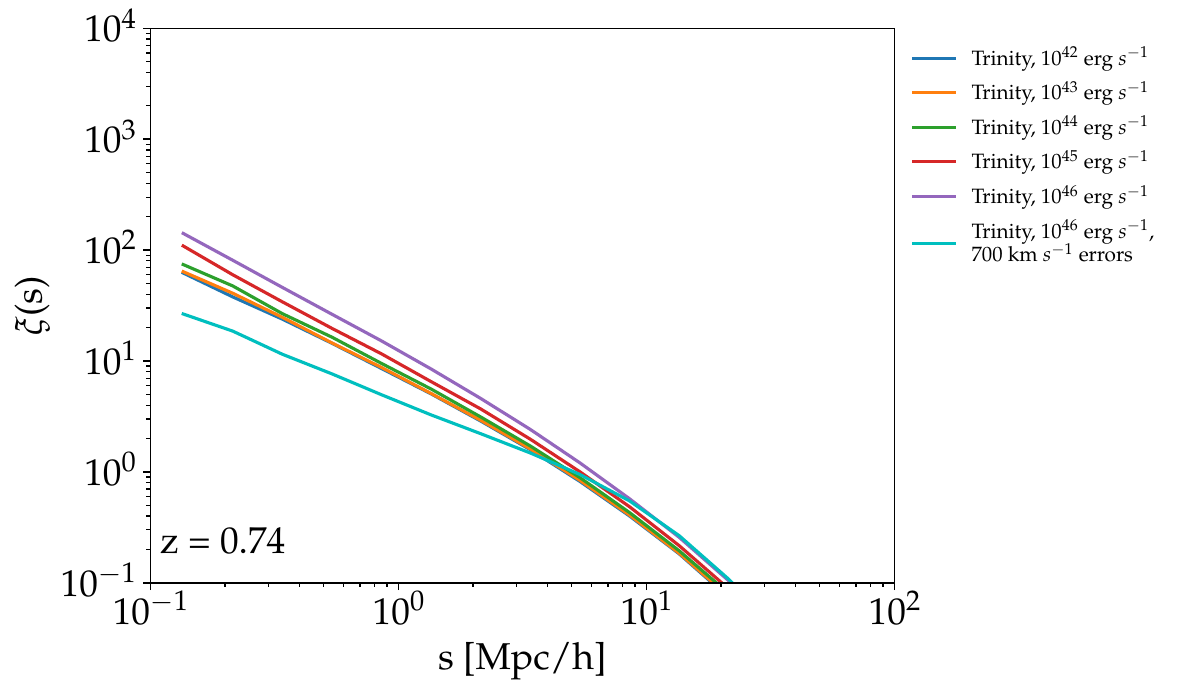}\\
	\includegraphics[width = 0.5\columnwidth]{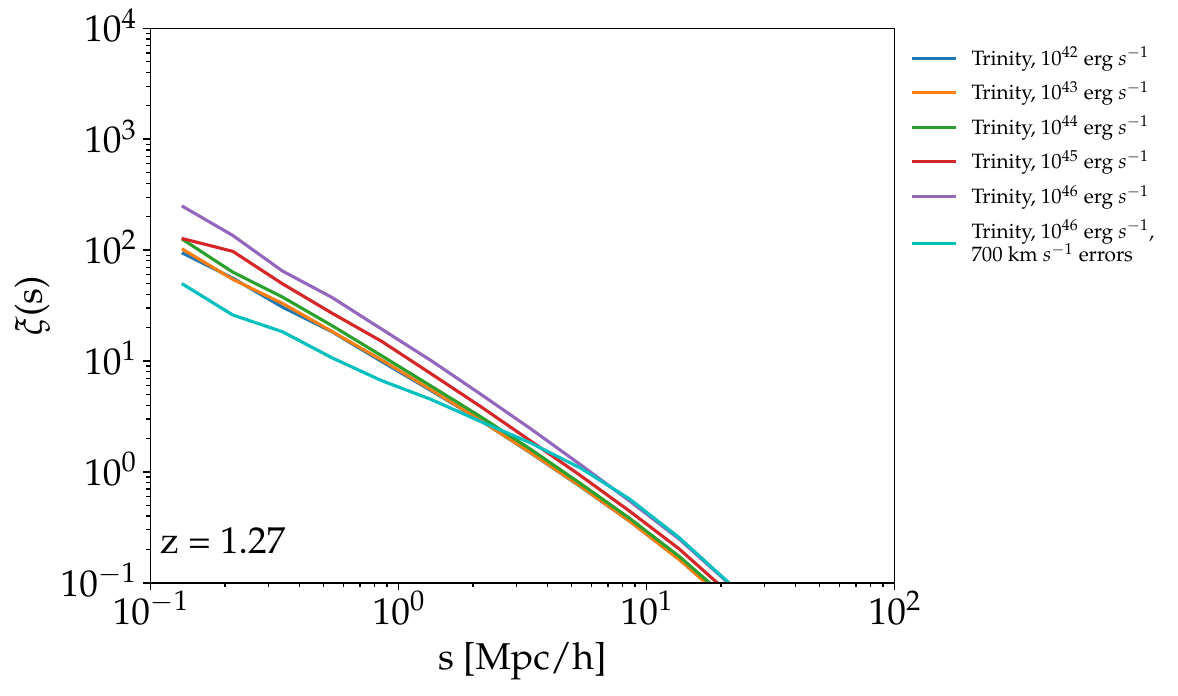} & 
	\includegraphics[width = 0.5\columnwidth]{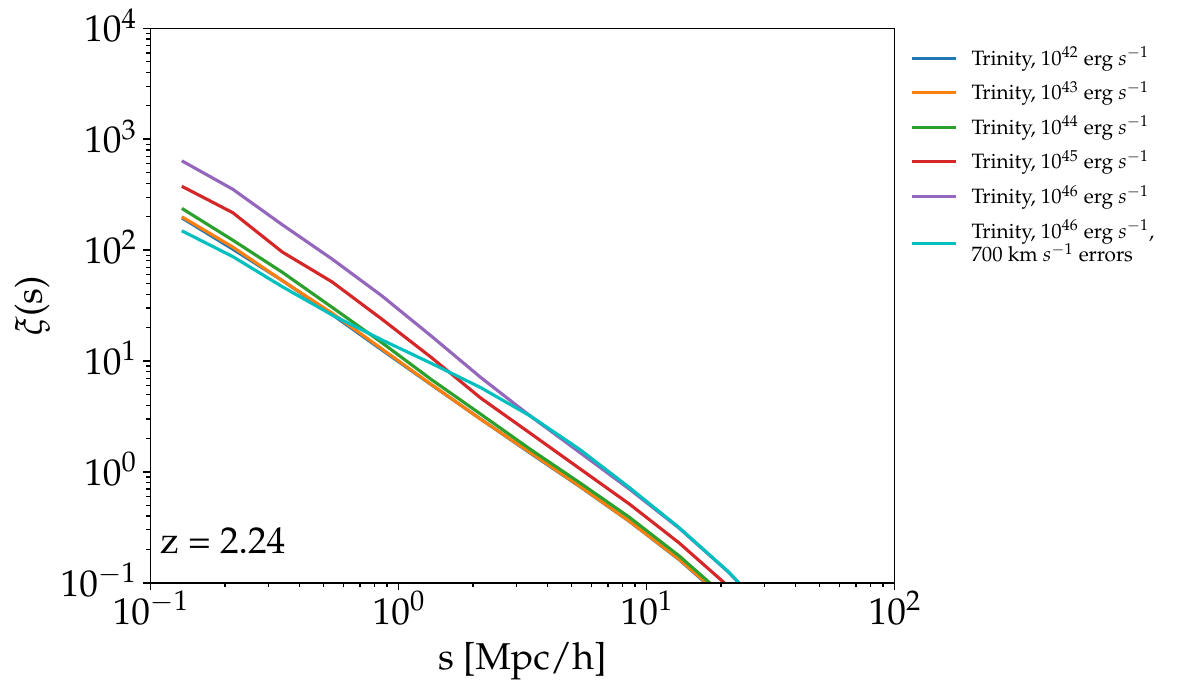}
	\end{tabular}
    \caption{The 3D redshift-space correlation function, $\xi_s$, for \textsc{TRINITY} from 10$^{42}$ - 10$^{46}$ erg s$^{-1}$. Each luminosity bin is represented by a different color.}
\end{figure*}

\end{document}